\documentclass[12pt]{article}
\usepackage[margin=1in]{geometry}
\usepackage{lineno}
\usepackage[colorlinks=true,citecolor=blue,pdfcreator={},pdfproducer={},linkcolor=blue]{hyperref}
\modulolinenumbers[1]
\usepackage{amsmath}
\usepackage{amsfonts}
\usepackage{bm}
\usepackage{graphicx}
\allowdisplaybreaks
\usepackage{amsbsy}
\usepackage{amssymb}
\usepackage[ruled]{algorithm2e}
\usepackage{natbib}
\usepackage[export]{adjustbox}

\usepackage{chngcntr}

\usepackage{subfig}

\usepackage[ruled]{algorithm2e}

\usepackage{indentfirst}
\usepackage{enumitem}
\setlist{nolistsep}

\newcommand*\patchAmsMathEnvironmentForLineno[1]{%
\expandafter\let\csname old#1\expandafter\endcsname\csname #1\endcsname
\expandafter\let\csname oldend#1\expandafter\endcsname\csname end#1\endcsname
\renewenvironment{#1}%
{\linenomath\csname old#1\endcsname}%
{\csname oldend#1\endcsname\endlinenomath}}%
\newcommand*\patchBothAmsMathEnvironmentsForLineno[1]{%
\patchAmsMathEnvironmentForLineno{#1}%
\patchAmsMathEnvironmentForLineno{#1*}}%
\AtBeginDocument{%
\patchBothAmsMathEnvironmentsForLineno{equation}%
\patchBothAmsMathEnvironmentsForLineno{align}%
\patchBothAmsMathEnvironmentsForLineno{flalign}%
\patchBothAmsMathEnvironmentsForLineno{alignat}%
\patchBothAmsMathEnvironmentsForLineno{gather}%
\patchBothAmsMathEnvironmentsForLineno{multline}%
}

\usepackage[nomarkers]{endfloat}

\DeclareGraphicsExtensions{.jpg}

\usepackage{xspace}

\usepackage{authblk}
\usepackage{times}

\title{\textbf{Three-dimensional seismic characterization and imaging of the Soda Lake geothermal field}}
\author[,1]{Kai Gao\thanks{Corresponding Author: kaigao@lanl.gov 
(K.~Gao); ljh@lanl.gov (L.~Huang)}}
\author[1]{Lianjie Huang}
\author[2]{Trenton Cladouhos}
\affil[1]{Los Alamos National Laboratory, Geophysics Group, MS~D446, Los Alamos, NM
	87545, USA}
\affil[2]{Cyrq Energy, Inc.,
4010 Stone Way North, Suite 400,
Seattle, WA 98103, USA}

\date{}
\providecommand{\keywords}[1]{\textbf{\textbf{Keywords:}} #1}

\begin{document}
\maketitle

\begin{abstract}
Accurate characterization of subsurface geophysical properties and 
detection of the fault system are essential for geothermal energy 
exploration and production. The Soda Lake geothermal field is in western Nevada with a complex fault system.  
Previous seismic characterization only produced a low-resolution, smooth 
velocity model along with a simple, conceptual fault model. Using 
optimized correlation-based full-waveform inversion, 
wavefield-separation-based reverse-time migration, and automatic fault 
detection techniques, we present 3D seismic characterization for the Soda 
Lake geothermal field using 3D surface seismic data acquired with 
Vibroseis sources. We obtain 3D high-resolution velocity, density, and 
acoustic impedance models, 3D seismic images with different grid 
spacings, and a high-resolution fault system. Consistency check between 
the constructed faults and currently active injection and production 
geothermal wells verifies that our seismic inversion and imaging results 
and detected faults are reliable.  These results can provide valuable 
information for optimizing well  placement and geothermal energy 
production at the Soda Lake geothermal field.
\end{abstract}


\section{Introduction}

\begin{figure}
	\centering
        \includegraphics[width=0.9\textwidth]{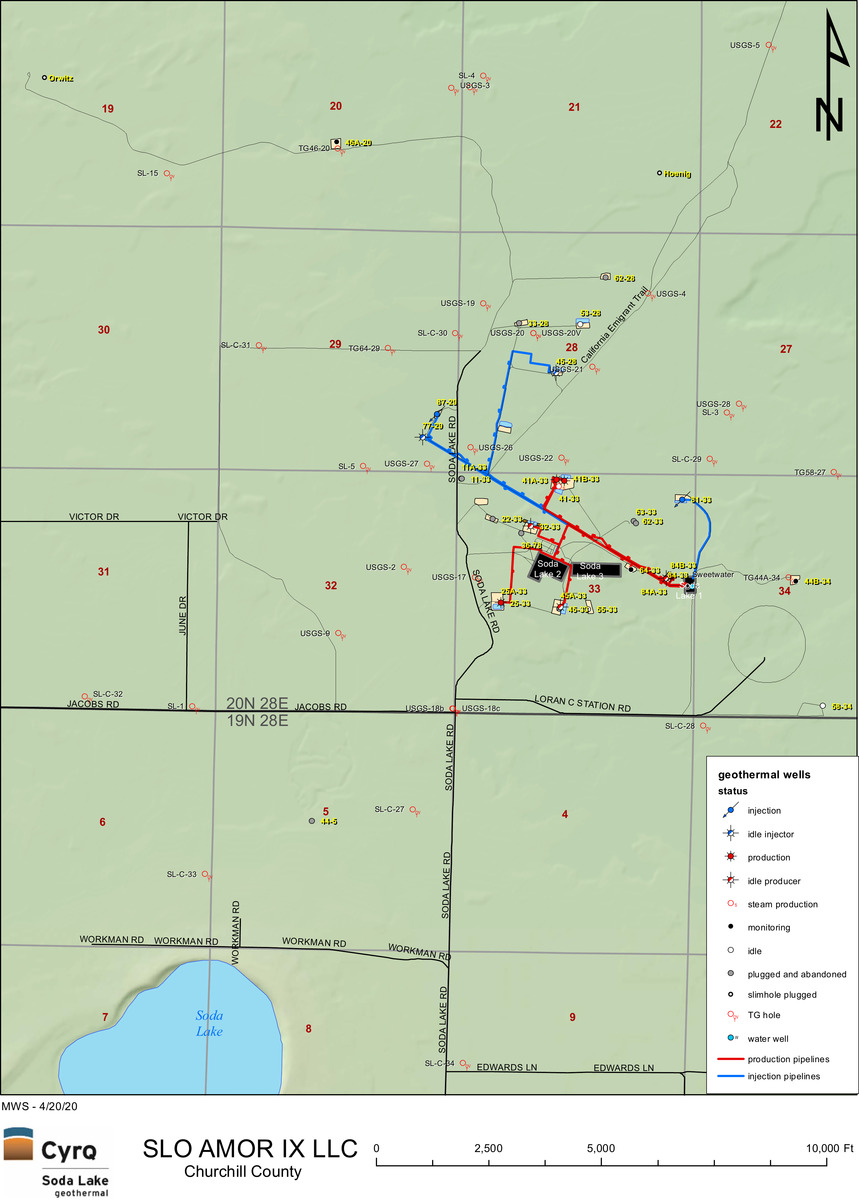}
        \caption{Location of the Soda Lake geothermal field in the 
        western Nevada.}
	\label{fig:map}
\end{figure}

Geothermal energy is an increasingly important component in the renewable 
energy sector of the United States. In the continental U.S., western 
states and Hawaii host most of the geothermal energy resources. The Soda Lake geothermal field is owned and operated by Cyrq Energy, Inc., and is located eight 
miles north-west of Fallon, Nevada, one of the most geothermal-rich 
states. Figure~\ref{fig:map} shows the location of the Soda Lake geothermal field. The Soda Lake geothermal plant is a binary geothermal 
electric-generating facility. Soda Lake I operated from 1987 through 2018, and Soda Lake II operated from 1991 through 2019. Figure~\ref{fig:map} 
shows the location of the Soda Lake geothermal field in the U.S.\ 
continent and its location in western Nevada. 

\begin{figure}
	\centering
        \includegraphics[width=\textwidth]{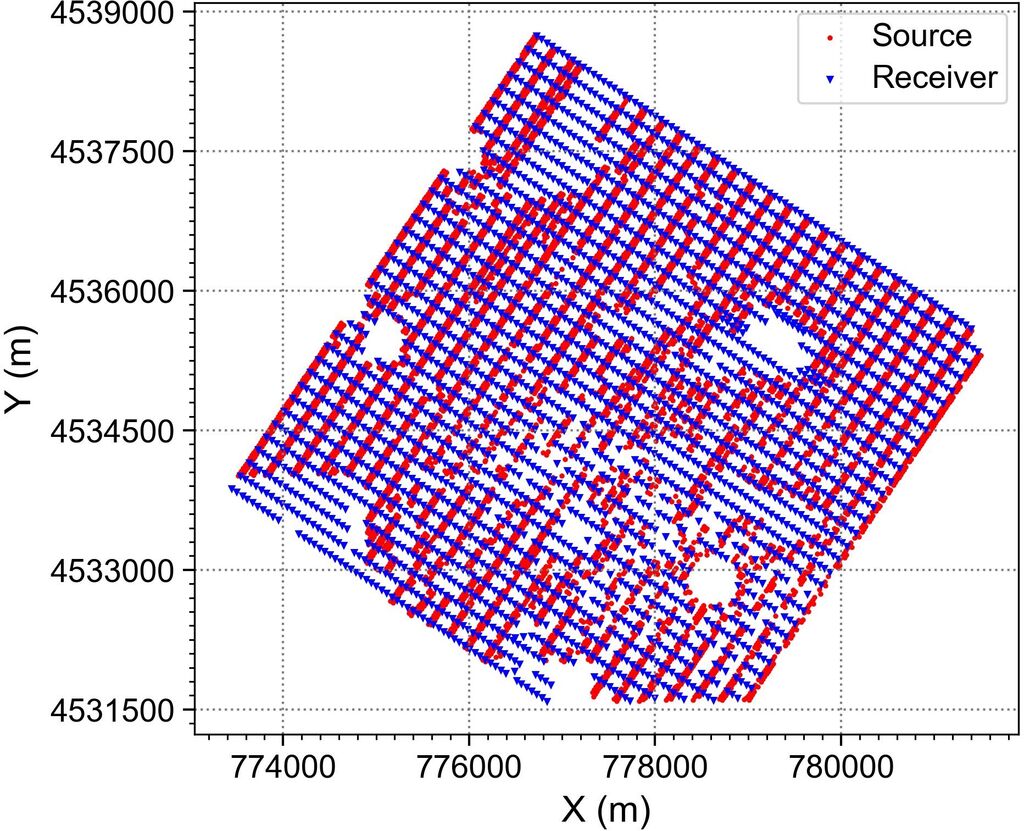}
        \caption{Source and receiver distributions of the 3D surface 
        seismic survey at the Soda Lake geothermal field. The survey 
     contains 8,321 common-shot gathers. Each common-shot gather has 
  a coverage area of up to approximately $3\times 3$~km$^2$.}
	\label{fig:geometry}
\end{figure}

In this study, we aim to provide a high-resolution, reliable, 3D seismic 
characterization for the Soda Lake geothermal field using  3D surface 
seismic data acquired in 2009. Previous studies 
constructed a low-resolution velocity model for the Soda Lake geothermal 
field, and built a conceptual fault model based on geophysical imaging 
and geological analysis. Our goal through this study is to reveal the 
complexity of subsurface geophysical properties in this region with our 
recently developed seismic-waveform inversion algorithm, and to construct 
a high-resolution fault system using our advanced seismic migration  
imaging and fault construction algorithms in a deterministic and 
systematic manners. 

The 3D surface seismic data acquired at the Soda Lake geothermal field 
contain a total of 8,321 compressional-to-compressional (PP)-component 
common-shot gathers.  Figure~\ref{fig:geometry} shows the distributions 
of sources (red dots)  and receivers (blue triangles) in this 3D seismic 
survey.  Each common-shot gather covers a surface area of up to 
approximately $3\times 3$~km$^2$.  The entire survey has a clockwise 
azimuth angle of 33.97$^{\circ}$, measuring from the positive Y direction 
(i.e., the North direction).  There exist some ``holes'' in the source 
and receiver distributions where there are no data acquired. The source 
lines are along the NE-SW direction with a line interval of 235.5~m.  The 
receiver lines are along the NW-SE direction with a line interval of 
167.5~m.  The inline source interval is 33.5~m, and the inline receiver 
interval is 67~m.  The surface area covered by all the sources and 
receivers is approximately $6\times 6$~km$^2$.

To obtain high-resolution subsurface medium parameter models for the Soda 
Lake geothermal field, we perform full-waveform inversion (FWI) of the 
acquired 3D surface seismic data. FWI is a nonlinear inversion method for 
estimating subsurface medium properties by minimizing the difference 
between observed seismic waveforms and synthetic waveforms  
\cite[]{Tarantola_1984,Virieux_Operto_2009}. First developed in 1980s, 
FWI is becoming a standard seismic inversion tool in both theoretical 
research and industry, particularly for highly complex geology. Suffering 
from high nonlinearity and the cycle-skipping issue, conventional FWI 
usually fails to converge to correct results for field seismic data, even 
for complex synthetic models. We employ our recently developed FWI 
algorithm that uses a different waveform match criterion in our algorithm 
to mitigate these difficulties as much as possible, even though not 
completely. Accompanied with our advanced, parallel full-waveform 
inversion codes that run on Los Alamos National Laboratory's
super-computing platform, we are able to perform a 3D full-waveform 
inversion for the Soda Lake geothermal field to obtain a set of 
high-resolution subsurface medium parameter models up to 2.5~km in depth. 

A subsurface structural image is one of the most important products from 
seismic characterization, which provides essential information for 
geothermal well placement and production. Previous studies have produced 
a subsurface image using conventional ray-based Kirchhoff migration that 
is incapable of handling complex structures. To improve subsurface 
imaging, we apply reverse-time migration to the 3D surface seismic data 
using our FWI-inverted 3D velocity model. Reverse-time migration (RTM) is 
an advanced seismic imaging technique for imaging complex structures 
\cite[]{McMechan_1983,Chang_McMechan_1987}. Using the PP seismic data 
separated from the 3D multi-component seismic data, we perform 
a wavefield-separation-based low-artifact RTM \cite[]{Fei_etal_2015} to 
obtain a high-resolution subsurface structural image for the Soda Lake 
geothermal field. We subsequently perform a fault-enhancing processing on 
the 3D RTM image and an automatic fault detection to delineate faults 
from the 3D image volume.  

Through this comprehensive seismic characterization, we achieve a set of 
high-resolution and reliable 3D subsurface medium parameter models, a 3D 
structural image, and a 3D fault system for the Soda Lake geothermal 
field.  Our inversion and imaging results reveal a complex fault system 
unravelling many faults that are not in the previous fault model. We find 
that current geothermal production wells either penetrate through the 
faults we detect from the image volume, or are fairly close to these 
faults. The consistency between the faults and currently active 
injection/production wells validates the accuracy and reliability of our 
inversion and imaging results and detected faults. Our high-resolution 3D 
subsurface medium property models and 3D fault system image can provide 
valuable information for optimizing well placement and geothermal energy 
production at the Soda Lake geothermal field.

Our paper is organized as follows: In the Methodology section, we briefly 
describe our seismic inversion and imaging methods applied to the 3D 
surface seismic data from the Soda Lake geothermal field. In the Results 
section, we present and analyze our seismic inversion and imaging 
results.  Particularly, we perform a check on accuracy and reliability of 
our imaging and fault detection results by plotting currently active 
geothermal injection and production wells in the 3D space.  We summarize 
our findings in the Conclusions section. 

\section{Methodology}

The seismic data we use in this study are pre-processed by Geokinetics, 
an industrial data processing company. Their data processing procedure 
includes random noise attenuation, groundroll attenuation, 
surface-consistent deconvolution and amplitude correction, spherical 
expanding amplitude compensation, and optionally automatic gain control 
or time-variant spectral whitening, etc.  Considering the complexity of 
3D seismic data processing for such a large dataset, we perform no 
additional data processing except frequency-domain filtering and 
offset-based data selection based on our needs for our 3D seismic 
inversion and imaging algorithms briefly described in this section. 

\subsection{Full-waveform inversion}

\begin{figure}
	\centering
	\includegraphics[width=\textwidth]{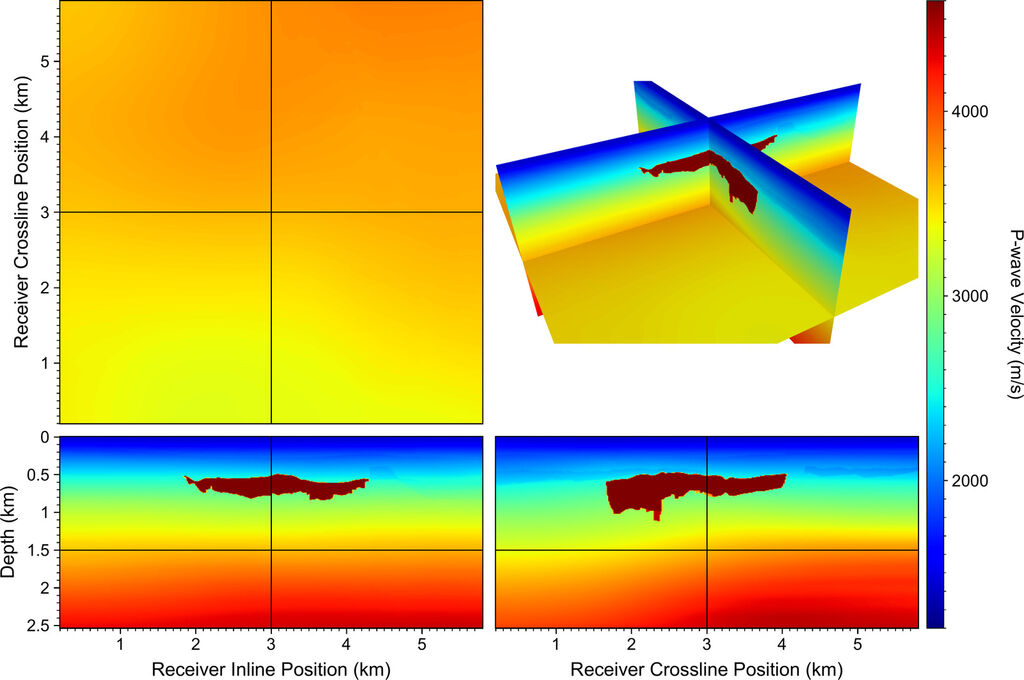}
        \caption{Slices and a 3D view of the initial P-wave velocity 
        model produced by Geokinetics at slicing position 1.}
	\label{fig:vpinit_1}
\end{figure}

\begin{figure}
	\centering
	\includegraphics[width=\textwidth]{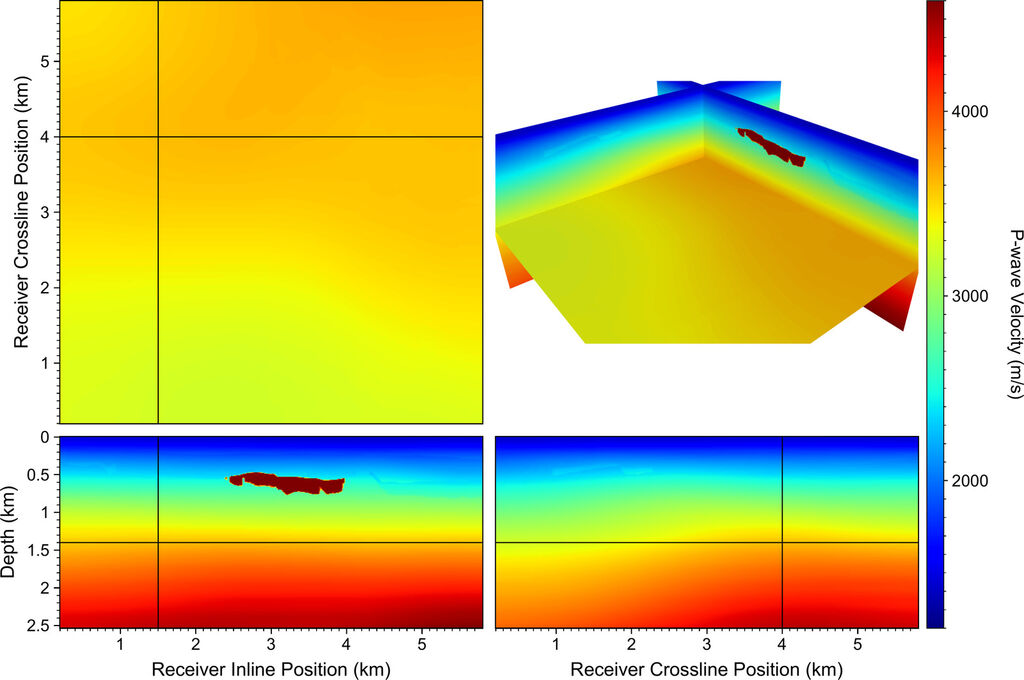}
        \caption{Slices and a 3D view of the initial P-wave velocity 
        model produced by Geokinetics at slicing position 2.}
	\label{fig:vpinit_2}
\end{figure}

\begin{figure}
	\centering
	\includegraphics[width=\textwidth]{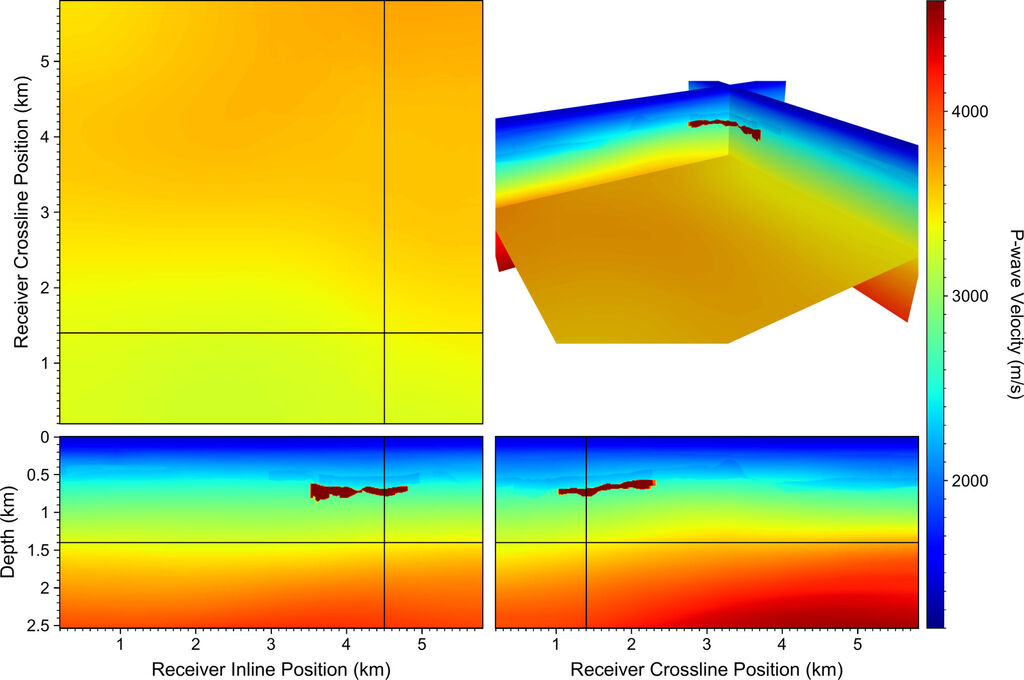}
        \caption{Slices and a 3D view of the initial P-wave velocity 
        model produced by Geokinetics at slicing position 3.}
	\label{fig:vpinit_3}
\end{figure}

Seismic processing by Geokinetics produced a smooth P-wave velocity model 
based on migration velocity analysis (MVA) and basalt body building.  
Figures~\ref{fig:vpinit_1}-\ref{fig:vpinit_3} display the smooth P-wave 
velocity model at three different slicing locations. The MVA also reveals 
a high-velocity basalt body at the center of the Soda Lake geothermal 
field, at the depth range of approximately 0.5 to 1~km. All the other 
regions of the initial velocity model is very smooth and contain almost 
no any high-wavenumber model perturbations that indicate either faults or 
sedimentary reflectors. 

We employ FWI to derive high-resolution subsurface medium property models 
for the Soda Lake geothermal field. FWI is a nonlinear inversion approach 
to estimating medium properties using both the amplitude and phase 
information of the full seismic wavefield 
\cite[]{Tarantola_1984,Plessix_2006,Virieux_Operto_2009}. In its simplest 
form, FWI is a $L_2$-norm nonlinear optimization problem:
\begin{equation}
\psi(\mathbf{m}) = \min_{\mathbf{m}} \frac{1}{2}||\mathbf{u} - \mathbf{d}||_2^2, 
\label{eq:fwi}
\end{equation}
where $\mathbf{m}$ is the model parameter, $\mathbf{d}$ is the observed 
seismic waveform, and $\mathbf{u}$ is the synthetic seismic waveform. FWI 
is applicable in either acoustic or elastic media. In our seismic 
characterization, we use its acoustic form and invert for both the P-wave 
velocity $V_p$ and the density $\rho$, i.e., $\mathbf{m} = (V_p, \rho)$. 
Inverting the density model in addition to the P-wave velocity model 
facilitates the FWI to achieve better amplitude match. Without including 
density inversion, some of the reflections caused by acoustic impedance 
contrasts would be attributed entirely to velocity contrasts of the 
model, leading to ``over-update'' of the P-wave velocity model. In this 
case, visually, the P-wave velocity would contain fairly obvious 
high-wavenumber perturbations that are geologically less plausible. 

FWI in the form of equation~\eqref{eq:fwi} is highly nonlinear and 
difficult to converge because of cycle-skipping issue, particularly for 
noisy field seismic data. To alleviate these difficulties, we adopt an 
optimized correlation misfit function in our seismic inversion  
\cite[]{Choi_Alkhalifah_2016}: \begin{equation}
\psi(\mathbf{m}) = \min_{\mathbf{m}} \frac{1}{2} ||\mathcal{C}_{\tau}(\mathbf{u}, \mathbf{d}) - \mathcal{C}_{\tau}(\mathbf{d}, \mathbf{d})||_2^2, 
\label{eq:fwicross}
\end{equation}
where $\mathcal{C}$ is the cross-correlation operation, and $\tau$ is the 
time lag of the cross-correlation. Our studies show that this 
correlation-based FWI misfit function usually leads to much better 
convergence and reliable results compared with conventional FWI with  
equation~\eqref{eq:fwi}. 

In FWI, the inversion gradients associated with the medium parameters are computed using the adjoint-state method \cite[]{Plessix_2006,Virieux_Operto_2009}. Our FWI is based on the following first-order form of the acoustic-wave equation:
\begin{align}
\frac{\partial p}{\partial t} + K \nabla\cdot \mathbf{v} &= f, \\
\rho \frac{\partial \mathbf{v}}{\partial t} + \nabla p &=0, 
\end{align}
where $K = \rho V_p^2$ is the bulk modulus of the medium, and $f$ is the source term. We invert for $V_p$ and $\rho$ simultaneously, therefore we compute the gradients associated $V_p$ and $\rho$ based on the wave equation system as
\begin{align}
\nabla_{V_p} \psi &= - \sum_{N_s, N_r} \int_{0}^{T_{\max}} \frac{1}{\rho V_p^3} \frac{\partial p}{\partial t} p^{\dagger} dt,  \\
\nabla_{\rho} \psi & = \sum_{N_s, N_r} \int_{0}^{T_{\max}} \frac{\partial 
\mathbf{v}}{\partial t} \mathbf{v}^{\dagger} dt,
\label{eq:fwi_grad}
\end{align}
where $p = p(\mathbf{x}, t)$ is the source pressure wavefield, $p^{\dagger}$ is the adjoint-state pressure wavefield, $\mathbf{v}$ and $\mathbf{v}^{\dagger}$ are the source and adjoint particle velocity wavefields, respectively, $N_s$ and $N_r$ are the numbers of sources and receivers, respectively, and $T_{\max}$ is the maximum propagation time of the wavefield. 

The adjoint-state wavefields $p^{\dagger}$ and $\mathbf{v}^{\dagger}$ are 
the solutions to the adjoint-state wave equation with the adjoint source 
being the external source term, solved in a reverse-time manner. The 
adjoint source is computed based on equation~\eqref{eq:fwicross}. One can 
refer \cite{Choi_Alkhalifah_2016} for the detailed expression of the 
adjoint source term.

\subsection{Reverse-time migration}

High-resolution subsurface structural images can reveal faults that are 
crucial for optimizing geothermal well placement and production.  
Conventional imaging such as ray-based Kirchhoff migration can produce  
satisfactory images for simple structures, but usually fails to provide 
clear and reliable images when the subsurface geological structures are 
complex, particularly for faulting geothermal fields. 

Reverse-time migration (RTM) is the industrial state-of-the-art imaging 
technique for imaging complex structures 
\cite[]{McMechan_1983,Chang_McMechan_1987}. RTM uses full wavefield to 
form subsurface images. In acoustic media, the PP image is formed by the 
zero time-lag cross-correlation between the source pressure wavefield and 
the receiver pressure wavefield. To properly attenuate low-wavenumber 
artifacts caused by high medium contrasts, as at the Soda Lake geothermal 
field containing a high-velocity basalt body, we employ the 
wavefield-separation-based RTM imaging condition \cite[]{Fei_etal_2015}:
\begin{equation}
I_{pp} = \sum_{N_s, N_r} \int_{0}^{T_{\max}} \left[p_s p_r - \mathcal{H}_z(p_s) \mathcal{H}_z(p_r) -p_s \mathcal{H}_z (q_r) - \mathcal{H}_z(p_s)  q_r \right]dt, 
\label{eq:ipp}
\end{equation}
where $N_s$ and $N_r$ are the numbers of source and receivers, 
respectively, $T_{\max}$ is maximum wavefield propagation time, and $p_s 
= p_s(\mathbf{x},t)$ and $p_r = p_r(\mathbf{x},t)$ are the source and 
receiver wavefields, respectively, with $\mathbf{x}$ being the spatial 
location and $t$ being the time. The wavefield $q_r = q_r(\mathbf{x},t) 
= g(\mathbf{x})*\mathcal{H}_t(d_0(t))$ is an auxiliary 
reverse-time-propagated wavefield with the temporal Hilbert transformed 
seismic data as the source term, $g = g(\mathbf{x})$ is Green's function, 
and $d_0 = d_0(t)$ is the recorded data at the receiver locations.  
$\mathcal{H}_z$ denotes the Hilbert transform in the depth direction. The 
auxiliary dataset $\mathcal{H}_t(d_0(t))$ is computed prior to RTM 
imaging.

\subsection{Fault detection}

Directly imaging faults with steep dips is a challenging task using 
seismic data \cite[e.g.,][]{Tan_Huang_2014}, which requires 
fault-generated seismic scattering wavefields well preserved in data.  
For seismic characterization at the Soda Lake geothermal field, we 
alternatively use post-imaging automatic fault detection method to 
delineate faults from the 3D seismic image volume produced using the 
aforementioned RTM algorithm. 

We delineate the subsurface fault system at the Soda Lake geothermal 
field using the optimal fault surface voting method 
\cite[]{Wu_Fomel_2018}. The method first automatically picks a set of 
sparse seed points from an initial input fault attribute image, and then  
uses them to construct the optimal surface patch based on global maximum 
attribute values. The method creates the final fault attribute map, such 
as the fault likelihood map, from the smoothed attribute maps based on 
collected accumulation scores. Finally, the method forms fault surfaces 
based on the computed fault strikes, dips and probabilities. One can 
refer to \cite{Wu_Fomel_2018} for algorithmic details. 

We then employ the detected faults to further enhance the RTM PP image 
with a fault-preserving, nonlinear anisotropic diffusion filtering method 
\cite[]{Fehmers_Hocker_2003,Wu_Guo_2018}. We process the PP image 
produced using RTM with the following nonlinear diffusion-type partial 
differential equation:
\begin{equation}
\frac{\partial I}{\partial t} = \nabla(\varepsilon \mathbf{D} \nabla I), 
\label{eq:andf}
\end{equation}
where $I=I(\mathbf{x})$ is the structural image, $\varepsilon=\varepsilon(\mathbf{x})$ is the spatial coherence information, e.g., the detected faults, and $\mathbf{D}=\mathbf{D}(\mathbf{x})$ is the spatially heterogeneous anisotropic diffusion tensor constructed from the structural tensor of the image $I(\mathbf{x})$. Equation~\eqref{eq:andf} can effectively suppress random noises, improve lateral continuity, while enhance faults of the image. One can refer to \cite{Fehmers_Hocker_2003} and \cite{Wu_Guo_2018} for algorithmic details. 

\section{Results}

\subsection{3D full-waveform inversion}

The initial MVA P-wave velocity model built by Geokinetics as shown in 
Figures~\ref{fig:vpinit_1}-\ref{fig:vpinit_3} satisfies some kinematic 
properties of the seismic data. To accommodate our FWI of the 3D surface 
seismic data from the Soda Lake geothermal field, we resample the initial 
velocity model with a 10~m regular grid spacing in three spatial 
directions.  This resampling results in a regular-grid initial P-wave 
velocity model of 603 sample points in both the X- and Y-directions, and 
255 sampling points in the depth direction. We build an initial density 
model using Gardner's rule \cite[]{Gardner_etal_1974} as $\rho = 310\times V_p^{0.25}$ with a unit 
of kg/m$^3$. 

Three-dimensional FWI is a computationally intensive inversion problem. 
Considering that the survey contains over 8,300 common-shot gathers, and 
the data have a relatively high signal-to-noise ratio after processing, 
we limit the dominant frequency of inversion to 10~Hz, and we display the 
inversion results at the 13th iteration.

\begin{figure}
	\centering
	\includegraphics[width=\textwidth]{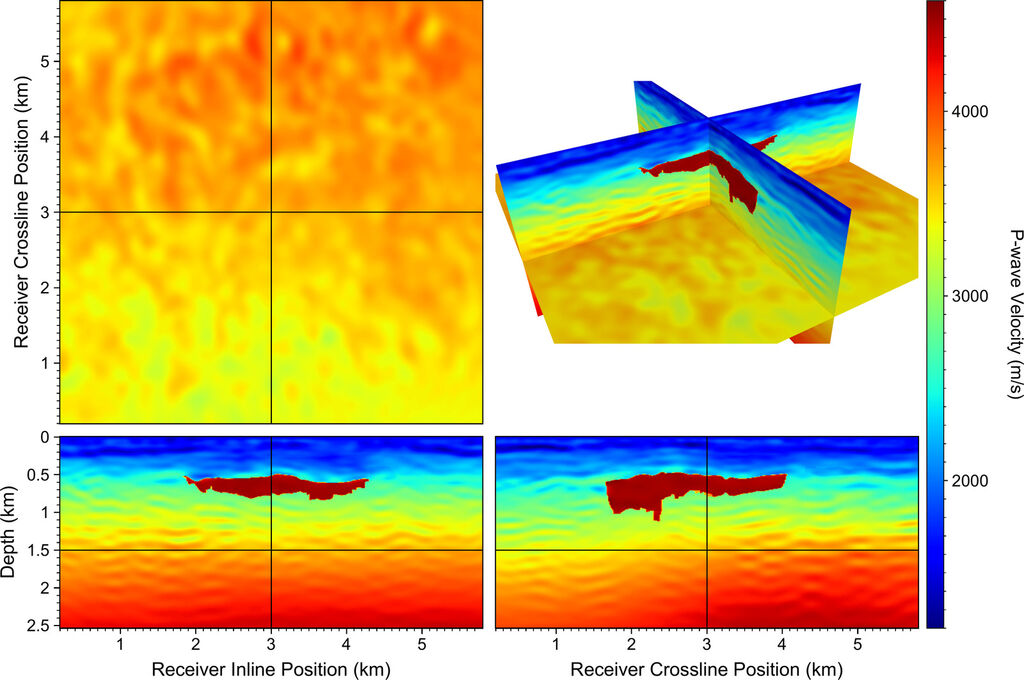}
        \caption{Slices and a 3D view of the updated P-wave velocity 
        model produced using our FWI at slicing position 1 .}
	\label{fig:vpfwi_1}
\end{figure}

\begin{figure}
	\centering
	\includegraphics[width=\textwidth]{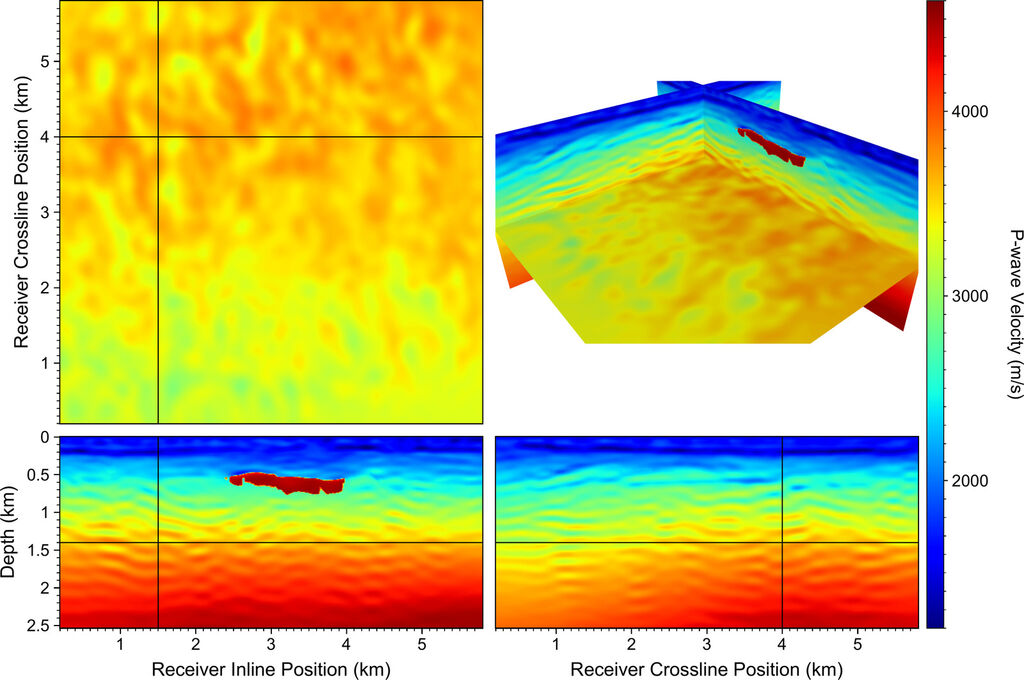}
        \caption{Slices and a 3D view of the updated P-wave velocity 
        model produced using our FWI at slicing position 2.}
	\label{fig:vpfwi_2}
\end{figure}

\begin{figure}
	\centering
	\includegraphics[width=\textwidth]{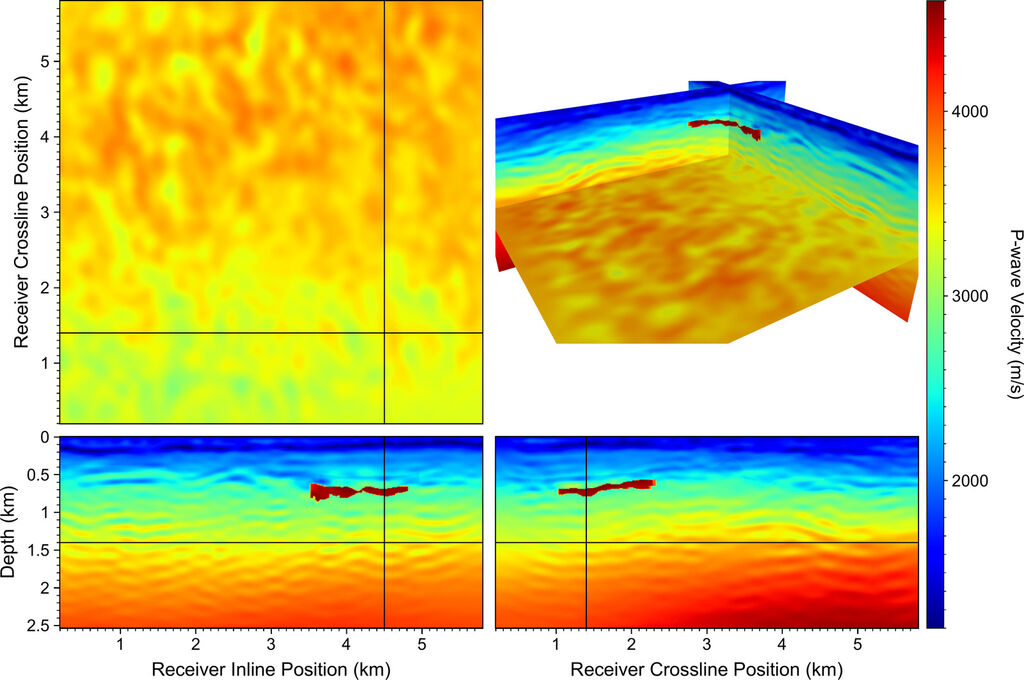}
        \caption{Slices and a 3D view of the updated P-wave velocity 
        model produced using our FWI at slicing position 3.}
	\label{fig:vpfwi_3}
\end{figure}

Figures~\ref{fig:vpfwi_1}-\ref{fig:vpfwi_3} display our FWI-inverted 
P-wave velocity model at three different slicing locations and view 
angles. We find apparent layer-structured model perturbations in both the 
shallow and deep regions compared with the initial smooth velocity model 
shown in Figures~\ref{fig:vpinit_1}-\ref{fig:vpinit_3}. Specifically, at 
the two slicing positions shown in Figures~\ref{fig:vpfwi_2} and 
\ref{fig:vpfwi_3}, we observe some faulting discontinuities cutting 
through the layered structures across the entire model, even at positions 
that are not beneath the high-contrasted basalt body. 

\begin{figure}
	\centering
	\includegraphics[width=\textwidth]{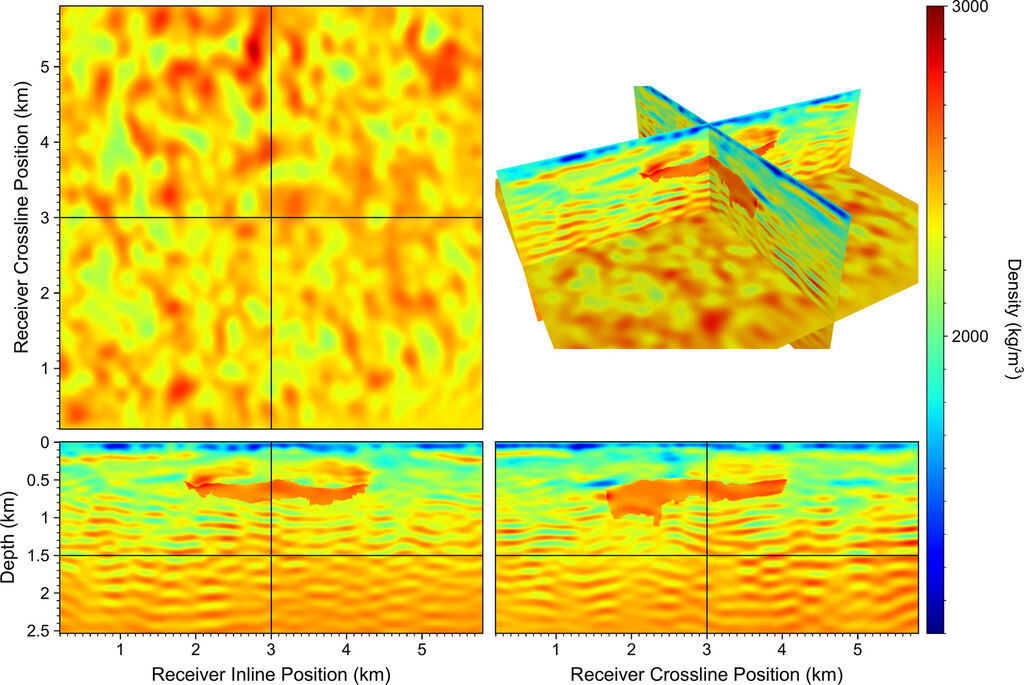}
                \caption{Slices and a 3D view of the updated density 
        model produced using our FWI at slicing position 1.}
	\label{fig:rhofwi_1}
\end{figure}

\begin{figure}
	\centering
	\includegraphics[width=\textwidth]{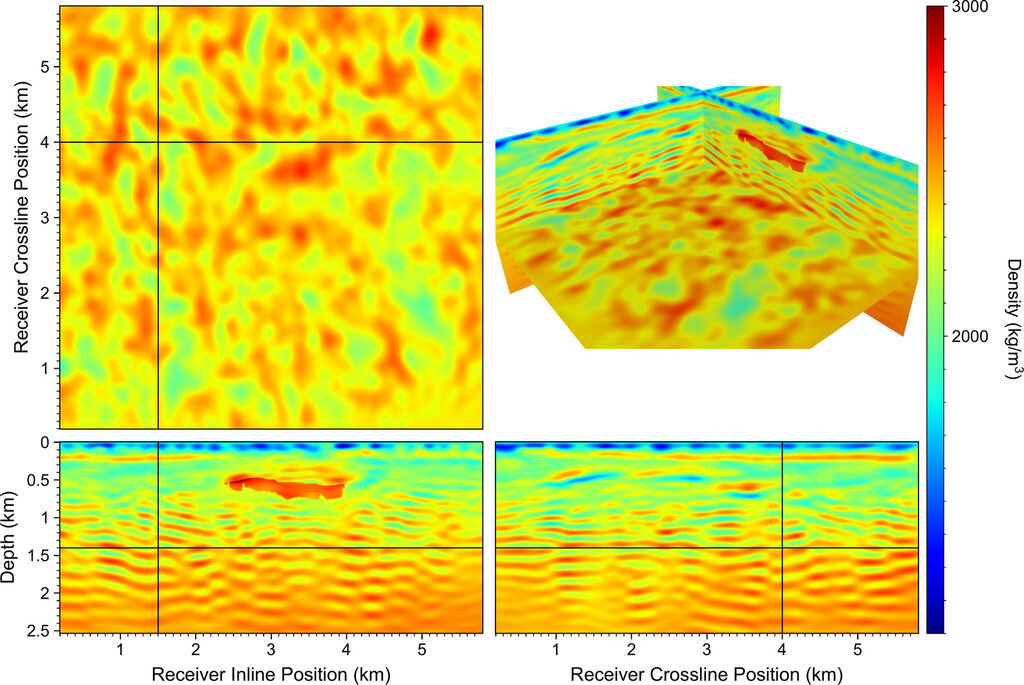}
                \caption{Slices and a 3D view of the updated density 
        model produced using our FWI at slicing position 2.}
	\label{fig:rhofwi_2}
\end{figure}

\begin{figure}
	\centering
	\includegraphics[width=\textwidth]{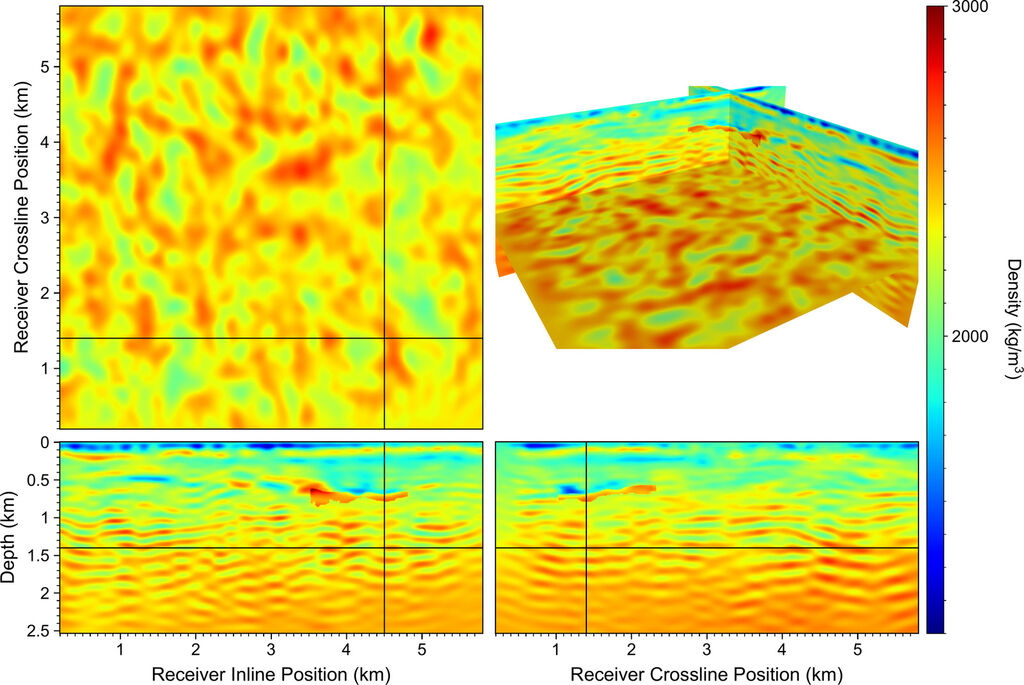}
                \caption{Slices and a 3D view of the updated density 
        model produced using our FWI at slicing position 3.}
	\label{fig:rhofwi_3}
\end{figure}

These layer-structured model perturbations and inferred faults are even 
more clear on our FWI-inverted density model shown in 
Figures~\ref{fig:rhofwi_1}-\ref{fig:rhofwi_3}. For example, in 
Figure~\ref{fig:rhofwi_2}, at inline receiver positions spanning from 
approximately 1.5~km to 5~km, we observe some clear faulting 
discontinuities that break the layers. Some of these faults are not 
beneath the basalt body, indicating that they are not artifacts caused by 
the high medium property contrast of the basalt body. We find similar 
faulting structures in Figure~\ref{fig:rhofwi_3}. 

The absolute values of the inverted density model might not be accurate 
out of two possible reasons. First, the seismic data input for the 
inversion is not truly acoustic, but is PP data separated from the 
acquired elastic multi-component data. Therefore, their amplitude might 
not be completely matched with the synthetic acoustic data. Second, we 
employ a cross-correlation type misfit function expressed in 
equation~\eqref{eq:fwicross}. This misfit function facilitates better 
convergence when using field seismic data, yet does not require a strict 
absolute amplitude match between the synthetic and the observed 
waveforms. Nevertheless, this inverted ``pseudo-density'' model, as 
described in our Methodology, prevents the inversion producing an 
over-updated velocity model. In this sense, we consider the inverted 
density model still has reasonable relative contrasts and accurate 
structures, although it may not be accurate in terms of absolute values. 

\begin{figure}
	\centering
	\includegraphics[width=\textwidth]{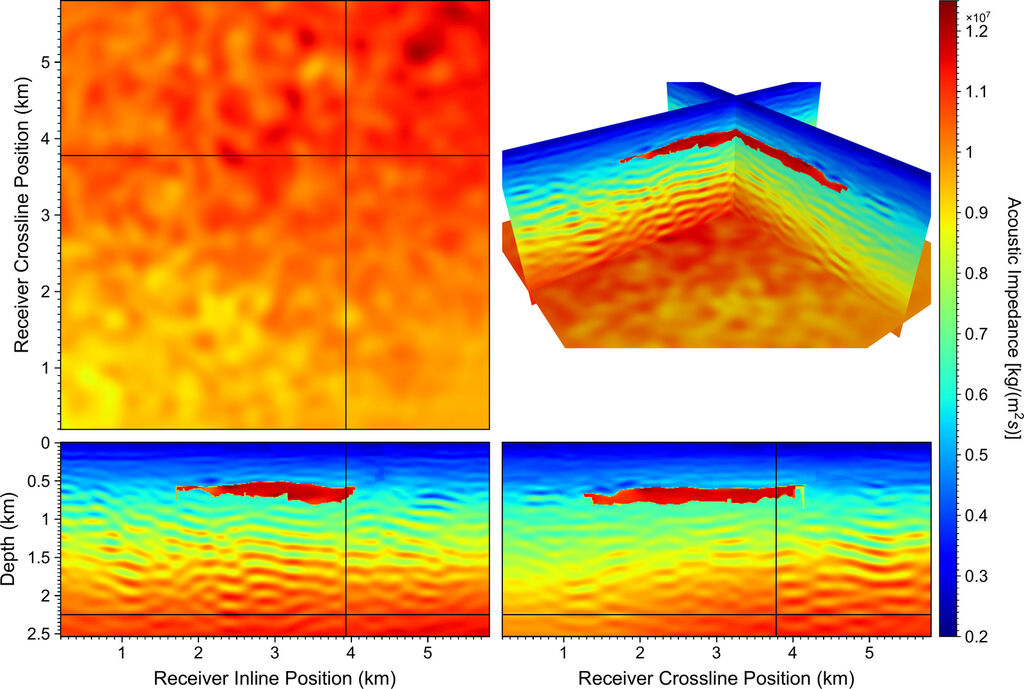}
        \caption{Slices and a 3D view of the updated acoustic impedance 
        model produced using our FWI.}
	\label{fig:aifwi_3}
\end{figure}

With both the FWI-updated P-wave velocity and density models, we compute 
their corresponding acoustic impedance model shown in 
Figure~\ref{fig:aifwi_3}, where we find clear discontinuities cutting 
through the layers. These features can be reasonably interpreted as 
faults in this region. In addition, we find clear layered structures in 
the acoustic impedance model, which generate reflections in the observed 
seismic data. These structures are obviously missing in the initial model 
shown in Figures~\ref{fig:vpinit_1}-\ref{fig:vpinit_3}. The spatial 
variations of the inverted models indicate the complexity of the 
subsurface structures at the Soda Lake geothermal field. 

\begin{figure}
	\centering
	\includegraphics[width=0.7\textwidth]{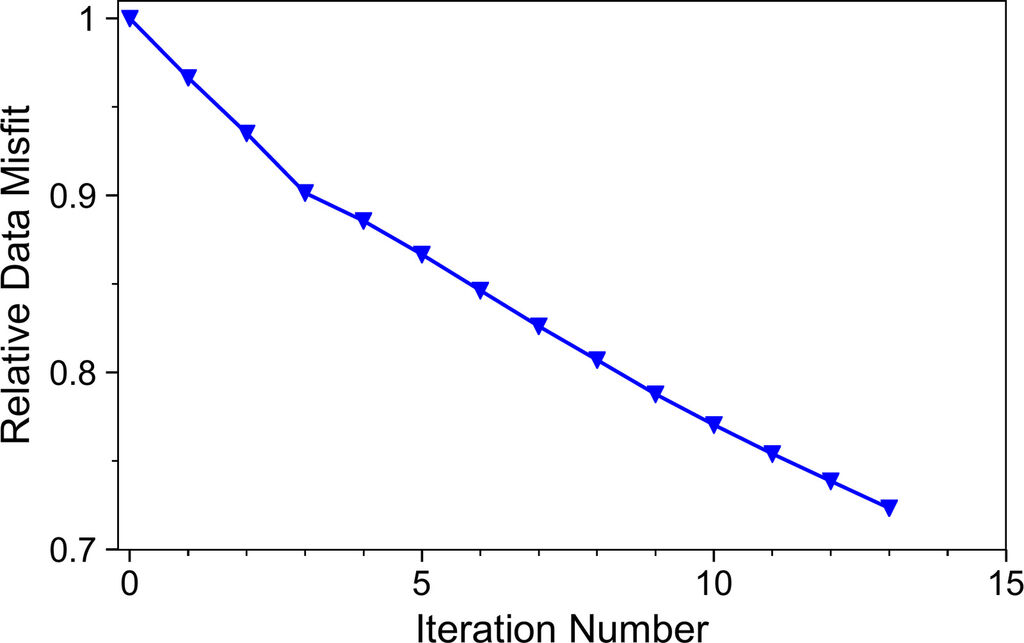}
	\caption{Data misfit convergence over a total of 13 iterations in our FWI.}
	\label{fig:misfit}
\end{figure}

Figure~\ref{fig:misfit} depicts the convergence of the relative data 
misfit of our FWI over a total of 13 iterations, showing an obvious 
misfit reduction even with the complexity of both the seismic data and 
the geological structures in this area. Additional updates may further 
reduce the data misfit.

\begin{figure}
	\centering
	\subfloat[]{\includegraphics[width=0.7\textwidth]{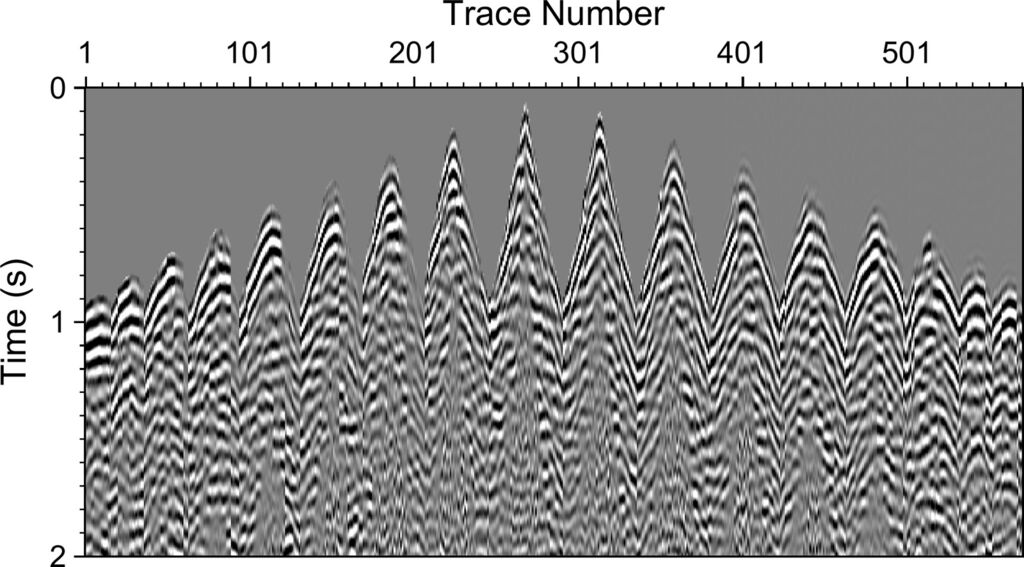}} \\
	\subfloat[]{\includegraphics[width=0.7\textwidth]{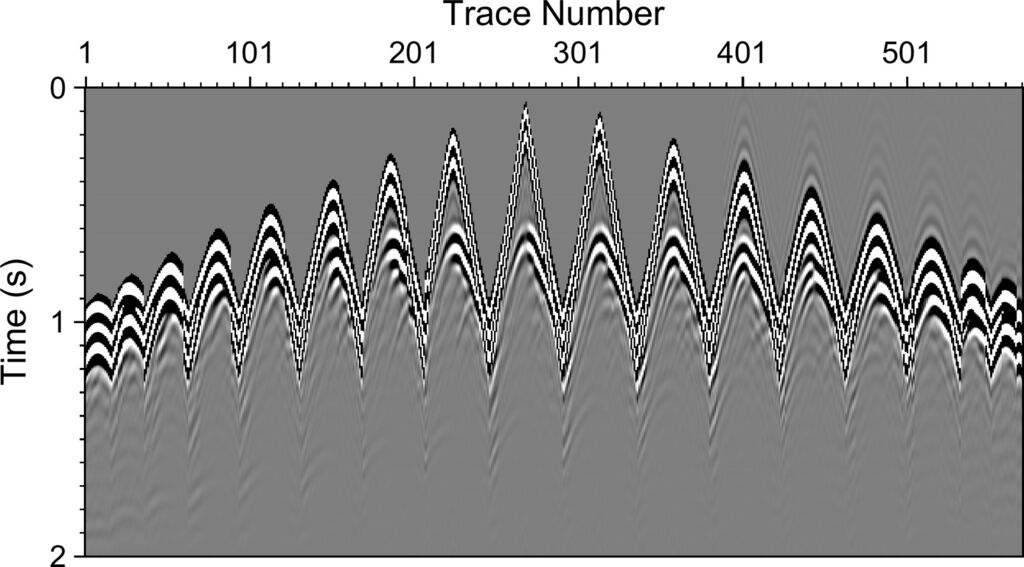}} \\
	\subfloat[]{\includegraphics[width=0.7\textwidth]{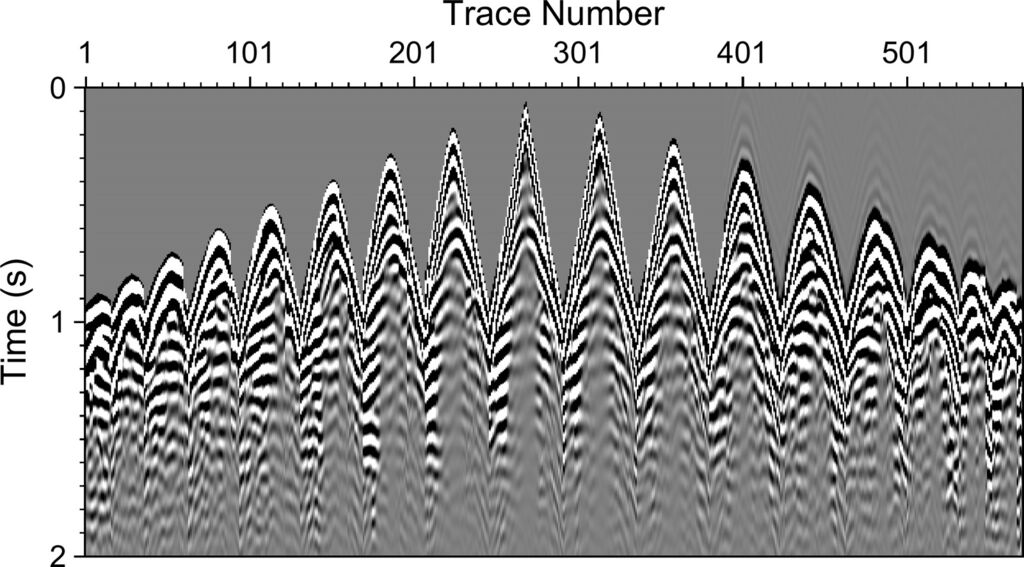}}
	\caption{Panels (a)-(c) show the observed data, the synthetic data in the initial model shown in Figure~\ref{fig:vpinit_1}, and the synthetic data in the FWI-updated model shown in Figure~\ref{fig:vpfwi_1}, respectively.}
	\label{fig:data}
\end{figure}

In Figure~\ref{fig:data}, we compare among the observed data 
(Figure~\ref{fig:data}a), the synthetic data in the initial model 
(Figure~\ref{fig:data}b) and the synthetic data in the FWI-inverted model 
(Figure~\ref{fig:data}c), for a randomly selected common-shot gather in 
the survey. The synthetic data in the initial model are obviously 
mismatched with the observed data.  Particularly, the reflection signals 
after approximately 1~s in the observed data are completely missing in 
the synthetic data. By contrast, the match between the synthetic data in 
the FWI-inverted model and the observed data is clearly improved.  
Seismic waveforms in Figure~\ref{fig:data}c before and after 1~s closely 
resemble those in the observed data in Figure~\ref{fig:data}a. Other 
common-shot gathers have also a similar data match improvement after our 
FWI. 

\subsection{3D reverse-time migration and fault detection}

We subsequently perform 3D reverse-time migration of the 3D surface 
seismic data from the Soda Lake geothermal field using the FWI-inverted 
velocity model. The center frequency of the source wavelet used in RTM is 
also 10~Hz, the same as that used in our FWI.

\begin{figure}
	\centering
	\includegraphics[width=\textwidth]{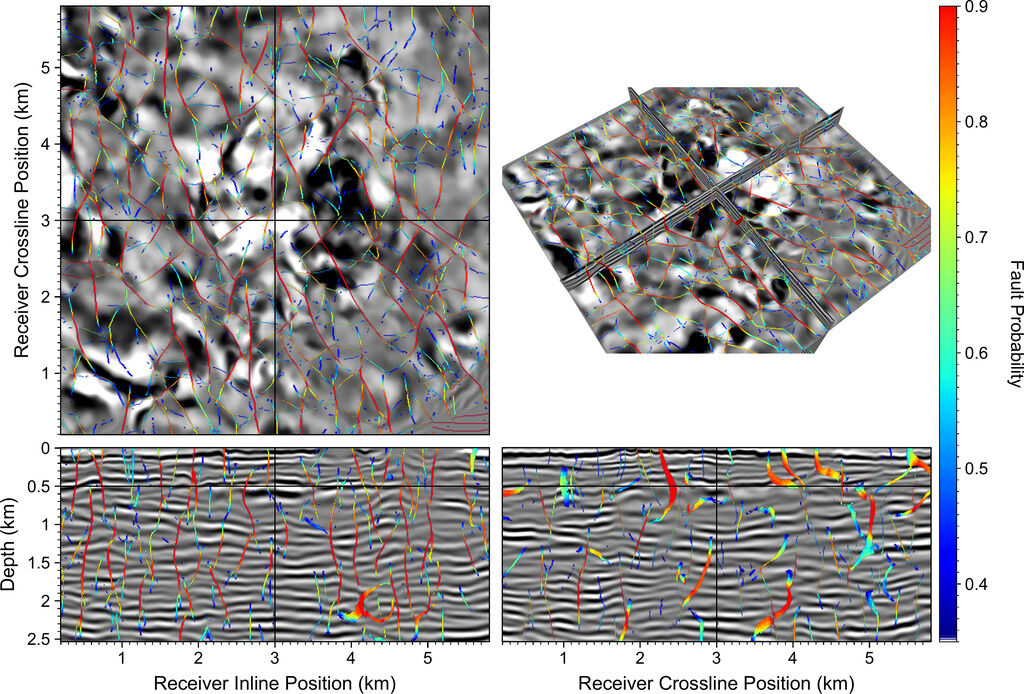}
        \caption{Slices and a 3D view of the structural image of the Soda 
        Lake geothermal field up to 2.5~km in depth produced with our 
     fault-enhancing RTM algorithm, superimposed with the detected faults 
  at slicing position 1.}
	\label{fig:ipp_lf0_1}
\end{figure}

\begin{figure}
	\centering
	\includegraphics[width=\textwidth]{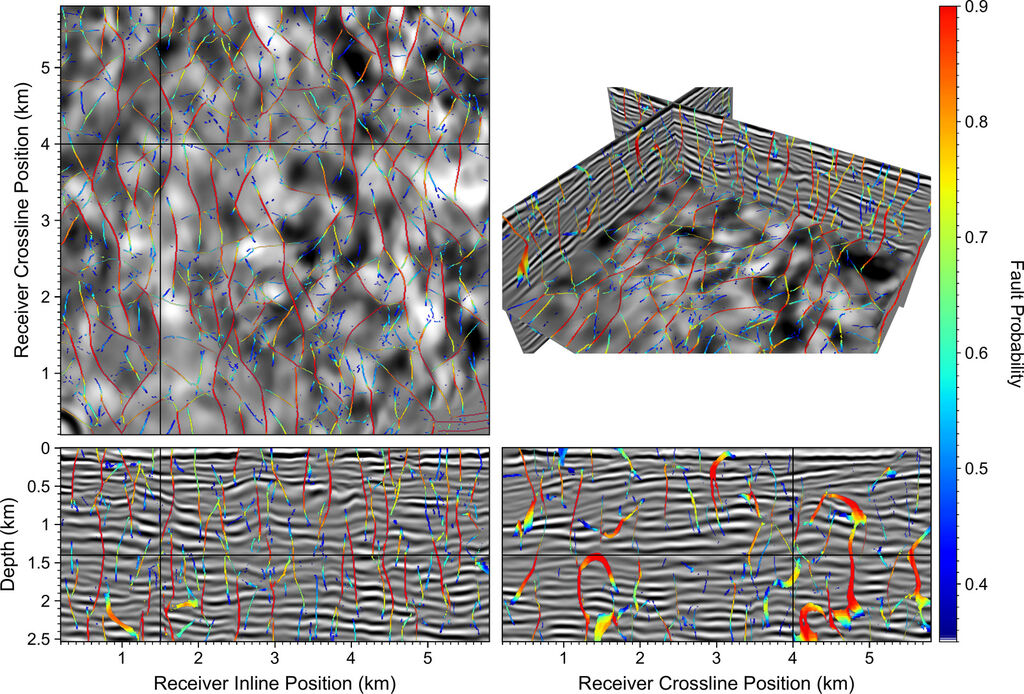}
        \caption{Slices and a 3D view of the structural image of the Soda 
        Lake geothermal field up to 2.5~km in depth produced with our 
     fault-enhancing RTM algorithm, superimposed with the detected faults 
  at slicing position 2.}
	\label{fig:ipp_lf0_2}
\end{figure}

\begin{figure}
	\centering
	\includegraphics[width=\textwidth]{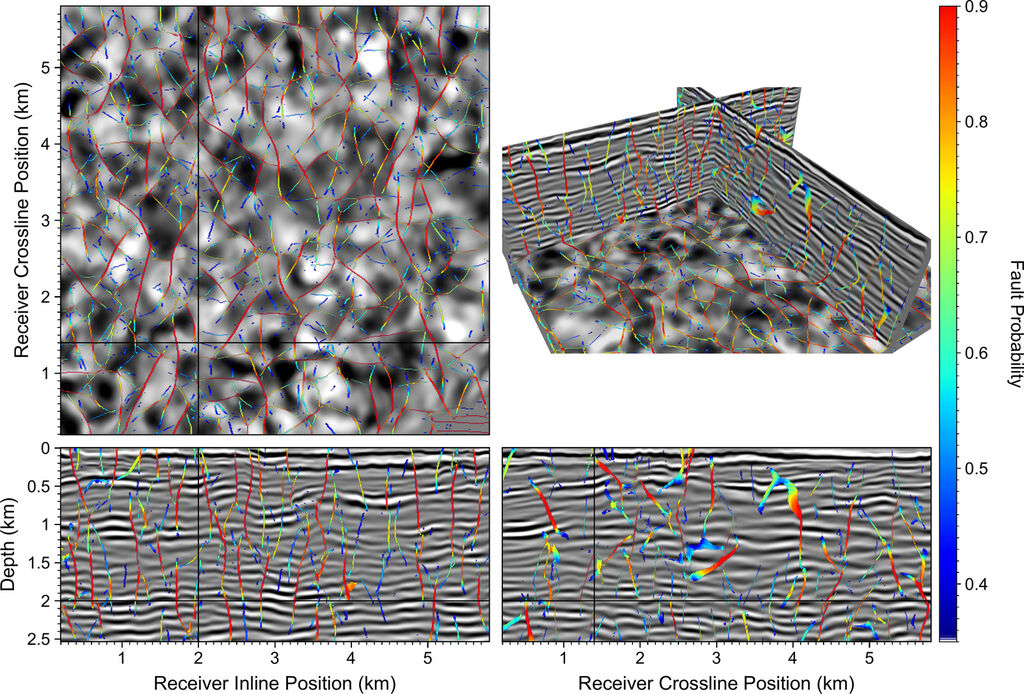}
	\caption{Slices and a 3D view of the structural image of the Soda 
        Lake geothermal field up to 2.5~km in depth produced with our 
     fault-enhancing RTM algorithm, superimposed with the detected faults 
  at slicing position 3.}
        \label{fig:ipp_lf0_3}
\end{figure}

Figures~\ref{fig:ipp_lf0_1}-\ref{fig:ipp_lf0_3} show the structural image 
up to 2.5~km in depth of the Soda Lake geothermal field at three 
different slicing positions. We also plot the detected faults along with 
the images, with colors of the faults representing fault probability.  
The horizontal slices on the top-left panels of 
Figures~\ref{fig:ipp_lf0_1}-\ref{fig:ipp_lf0_3} show that major faults in 
this area are along the diagonal direction of the geometry. Interpreted 
based on the map shown in Figure~\ref{fig:map}, these major faults are 
approximately along the north-south direction, with a small to moderate 
azimuth angle. On the vertical slices, we find the layers in this region 
are well imaged, even blow the high-contrasted basalt body. Faults in 
this region have steeply dipping angles, as indicated by the fault images 
in Figures~\ref{fig:ipp_lf0_1}-\ref{fig:ipp_lf0_3}. Some of the these 
faults penetrate through the basalt body, indicating that the faulting in 
this region occurred after the basalt body was formed in the geological 
history.

\begin{figure}
	\centering
	\subfloat[]{\includegraphics[width=0.6\textwidth]{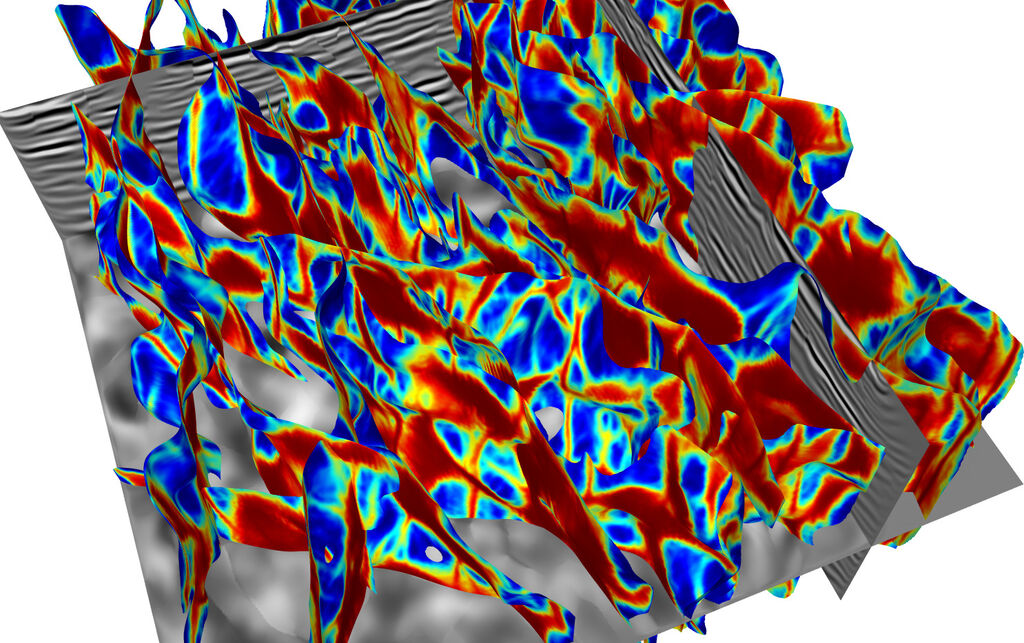}} \\
	\subfloat[]{\includegraphics[width=0.6\textwidth]{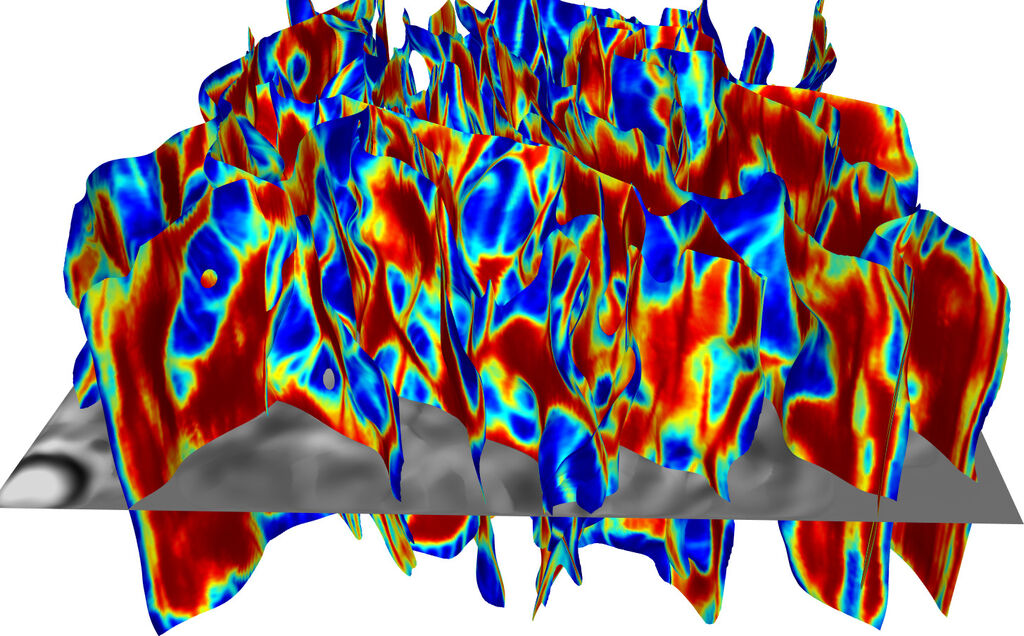}} \\
	\subfloat[]{\includegraphics[width=0.6\textwidth]{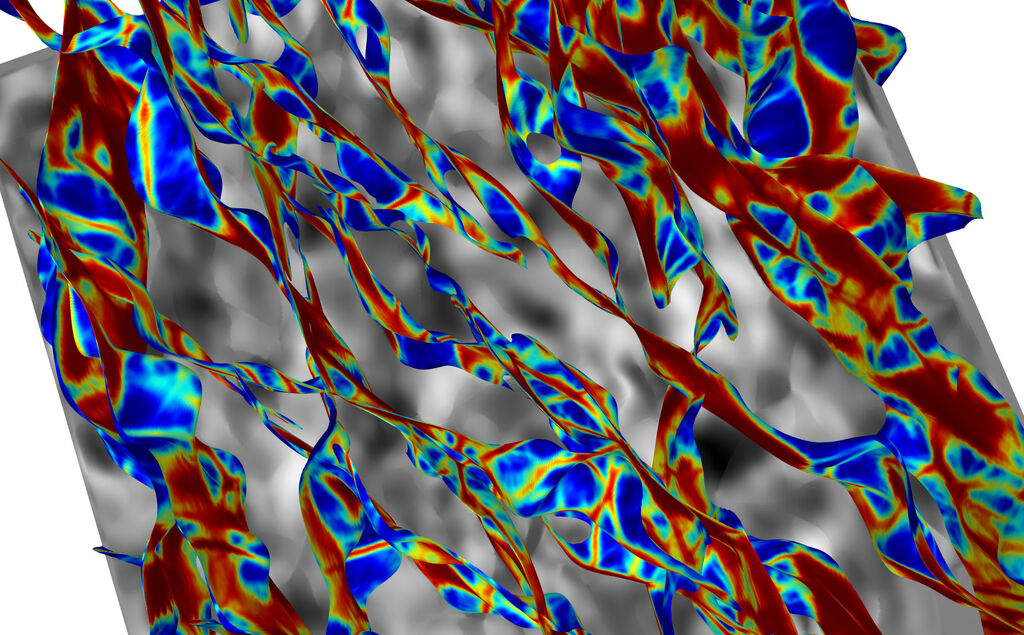}}
        \caption{Panels (a)-(c) show the constructed fault surfaces from 
        the RTM structural image shown in Figure~\ref{fig:ipp_lf0_1} at 
     three different view angles. Colors of the fault surfaces represent 
  the fault probability and are consistent with those in 
  Figures~\ref{fig:ipp_lf0_1}-\ref{fig:ipp_lf0_3}.}
	\label{fig:fault_lf0}
\end{figure}

Figure~\ref{fig:fault_lf0} shows the detected faults from the structural 
image shown in Figure~\ref{fig:ipp_lf0_1} at three different view angles. 
Consistent with those shown in 
Figures~\ref{fig:ipp_lf0_1}-\ref{fig:ipp_lf0_3}, the colors of the fault 
surfaces represent the fault probability. Although the fault probability 
is not full for every spatial point on the fault surfaces, most fault 
surfaces have moderate to high probability, indicating the detected 
faults are reliable.

\begin{figure}
	\centering
	\subfloat[]{\includegraphics[width=0.45\textwidth]{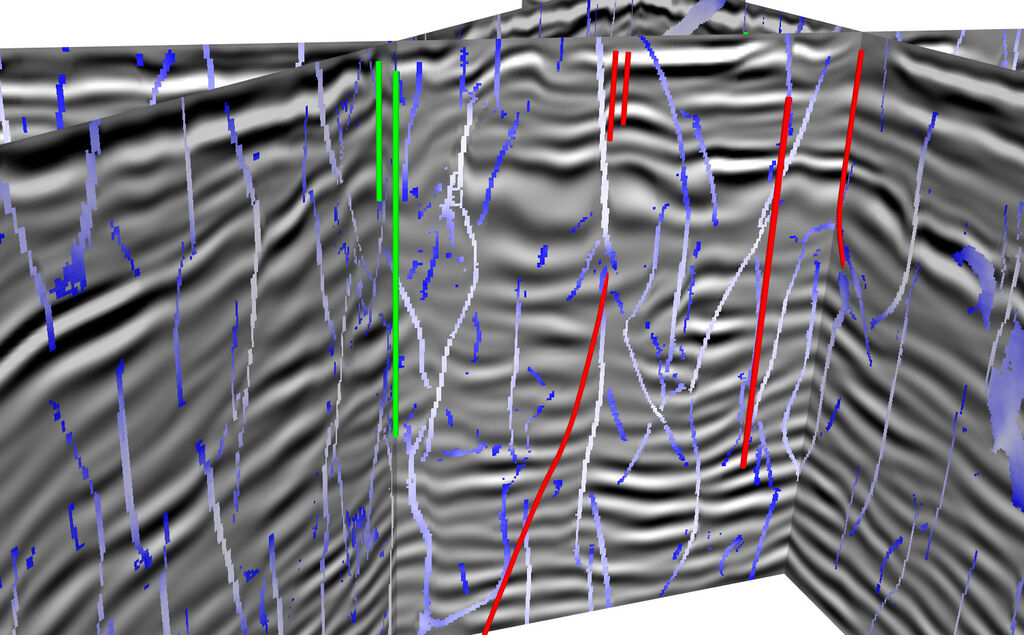}} \quad
	\subfloat[]{\includegraphics[width=0.45\textwidth]{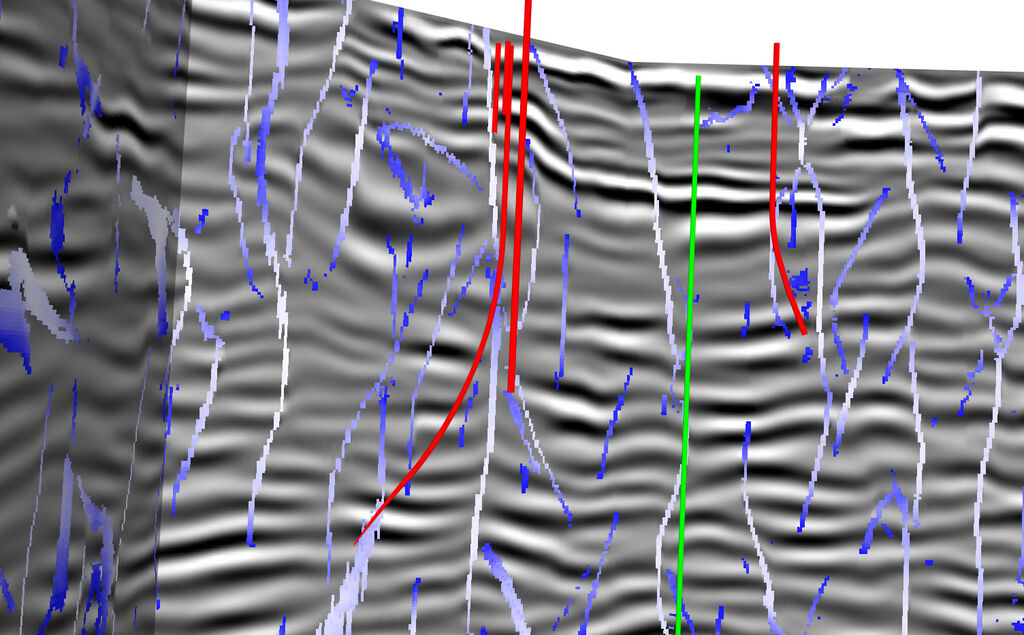}} \\
	\subfloat[]{\includegraphics[width=0.45\textwidth]{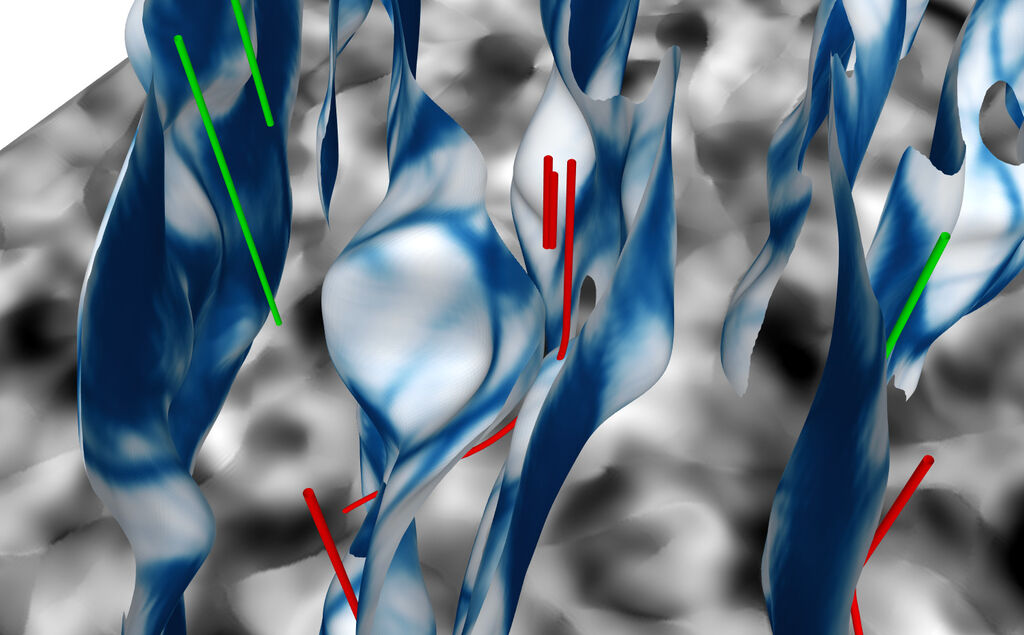}} \quad
	\subfloat[]{\includegraphics[width=0.45\textwidth]{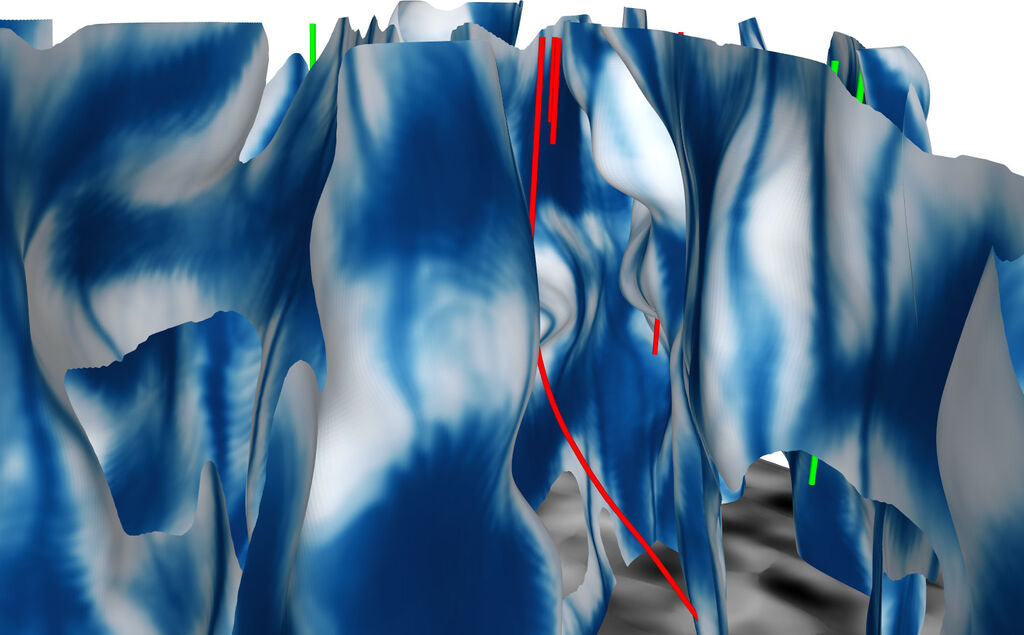}} \\
	\subfloat[]{\includegraphics[width=0.45\textwidth]{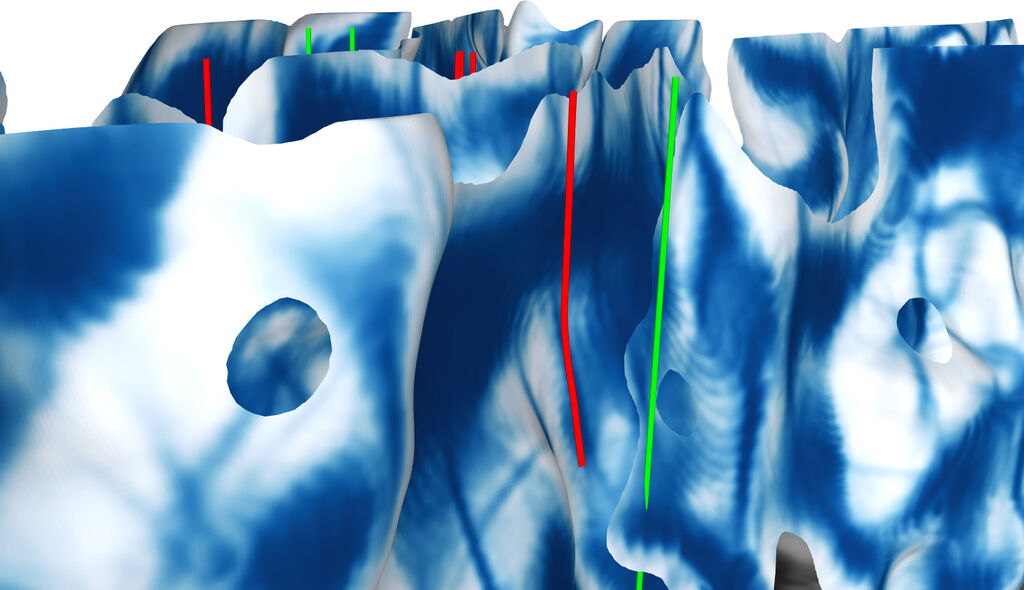}} \quad
	\subfloat[]{\includegraphics[width=0.45\textwidth]{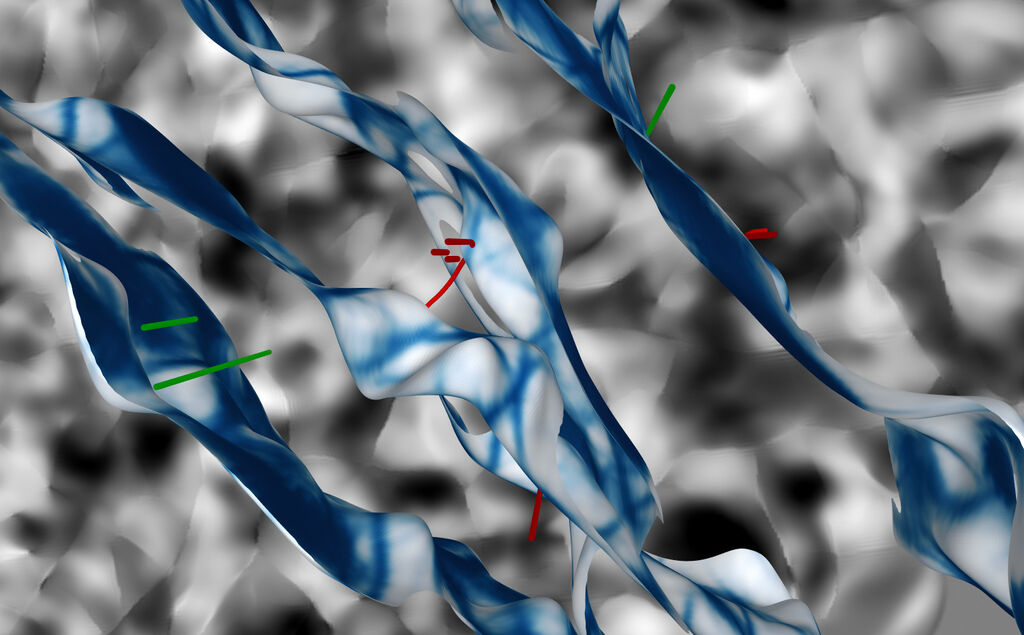}} \\
        \caption{Panels (a)-(b) show the detected faults for the RTM 
        structural image shown in Figure~\ref{fig:ipp_lf0_1} at two 
        different view angles, along with active injection (green) and 
     production (red) wells at the Soda Lake geothermal field. Panels 
  (c)-(f) show the constructed fault surfaces with currently active 
  injection and production wells in this geothermal field.}
	\label{fig:well_fault_lf0}
\end{figure}

There are several currently active injection and production wells at the 
Soda Lake geothermal field. To validate the accuracy and reliability of 
our imaging and fault detection results, we visualize the wells, the 
structural images, and the detected faults in their true 3D spatial 
positions, as displayed in Figure~\ref{fig:well_fault_lf0}.  
Figures~\ref{fig:well_fault_lf0}a and b show several image slices 
superimposed with the detected faults (in white-blue colors). We place 
the injection wells (green-colored tubes) and production wells 
(red-colored tubes) in the 3D space. We find that all the injection and 
production wells either run through the detected faults or are fairly 
close to one or two faults. For instance, in 
Figure~\ref{fig:well_fault_lf0}a, a curved production well in the center 
of the image penetrates exactly a major fault on the image.  The 
production well on the right of the figure also penetrates a location 
where two faults intersect. Figure~\ref{fig:well_fault_lf0}b shows that 
that the green-colored injection well on the right of the figure is 
located adjacent to a major fault, and part of the injection well 
overlies the fault path. 

We show the constructed fault surfaces in 
Figures~\ref{fig:well_fault_lf0}c-f, along with injection and production 
wells plotted in the same 3D scene. Clearly, all the production wells 
either penetrate one or two faults, or are very close to a fault surface.  
Some of the production wells penetrate through a same fault twice. The 
consistency between the currently active injection and production wells 
and our constructed faults from our 3D image volume demonstrates that our 
imaging and fault detection results are very close to the realistic 
geology in this region. The detected fault surfaces consist of a complex 
fault system in this region.

Our preceding results in this paper reveal complex subsurface structures 
and medium property variations up to 2.5~km at the Soda Lake geothermal 
field, with a spatial grid interval of 10~m along lateral and vertical 
directions.  We aim to reveal the complex fault system with a higher 
spatial resolution using a grid spacing of 6.7~m in the horizontal 
directions and 2.5~m in the depth direction, up to 1~km in depth for this 
area. We use a source wavelet with a center frequency of 30~Hz for RTM in 
this fine grid, enabling us to resolve fine layers for this geothermal 
field, and to obtain a high-resolution fault system. 

\begin{figure}
	\centering
	\includegraphics[width=\textwidth]{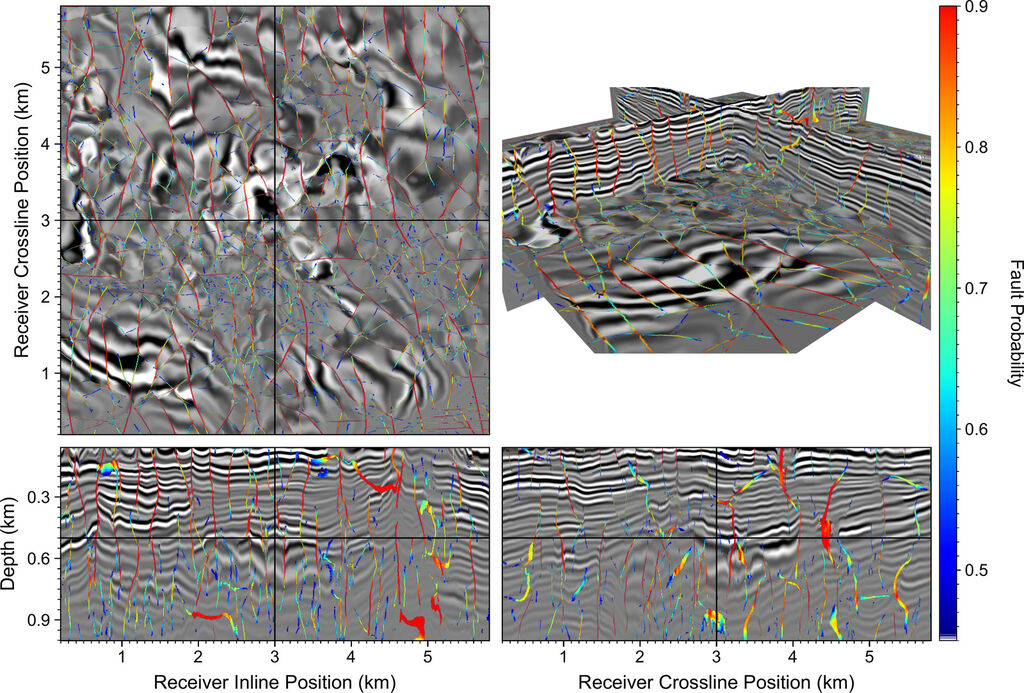}
        \caption{Slices and a 3D view of the structural image up to 1~km 
                in depth produced with our fault-enhancing RTM algorithm, 
             superimposed by the detected faults at slicing position 1.}
		\label{fig:ipp_hf0_1}
\end{figure}

\begin{figure}
	\centering
	\includegraphics[width=\textwidth]{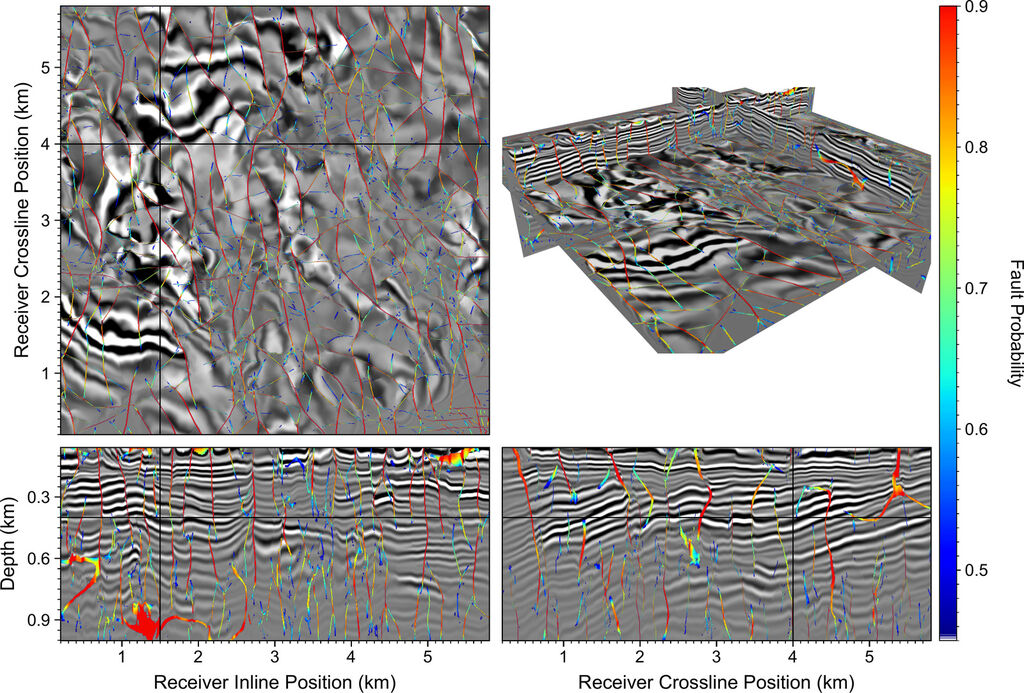}
	\caption{Slices and a 3D view of the structural image up to 1~km 
                in depth produced with our fault-enhancing RTM algorithm, 
                superimposed by the detected faults at slicing position 
             2.}
        \label{fig:ipp_hf0_2}
\end{figure}

\begin{figure}
	\centering
	\includegraphics[width=\textwidth]{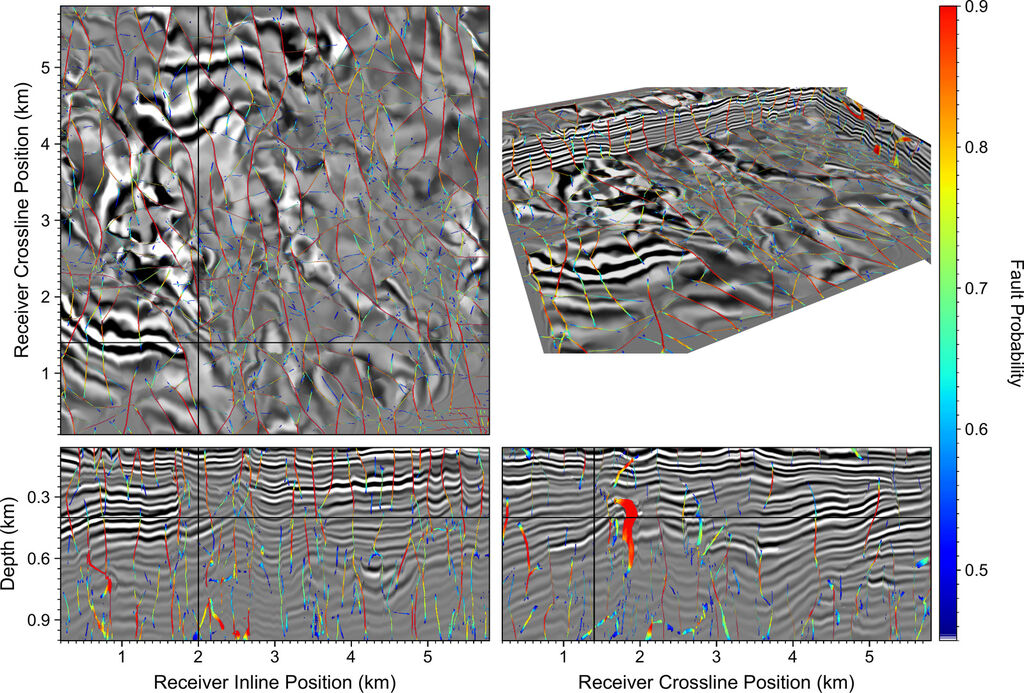}
	\caption{Slices and a 3D view of the structural image up to 1~km 
                in depth produced with our fault-enhancing RTM algorithm, 
                superimposed by the detected faults at slicing position 
             3.}
        \label{fig:ipp_hf0_3}
\end{figure}

Figures~\ref{fig:ipp_hf0_1}-\ref{fig:ipp_hf0_3} show the image volume 
with a horizontal grid spacing of 6.7~m and a vertical grid spacing of 
2.5~m, superimposed by the detected faults from this image volume.  We 
observe that the detected faults have similar strikes with those shown in 
Figures~\ref{fig:ipp_lf0_1}-\ref{fig:ipp_lf0_3}.  These high-resolution 
images and detected faults further verify the geological complexity of 
the near-surface region up to 1~km in depth. 

\begin{figure}
	\centering
	\subfloat[]{\includegraphics[width=0.6\textwidth]{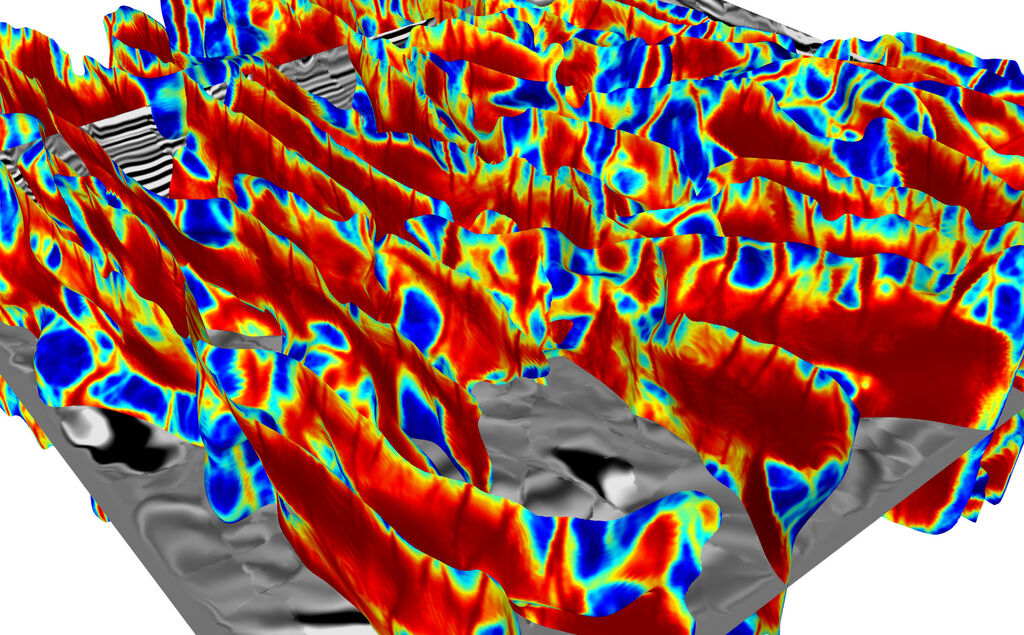}} \\
	\subfloat[]{\includegraphics[width=0.6\textwidth]{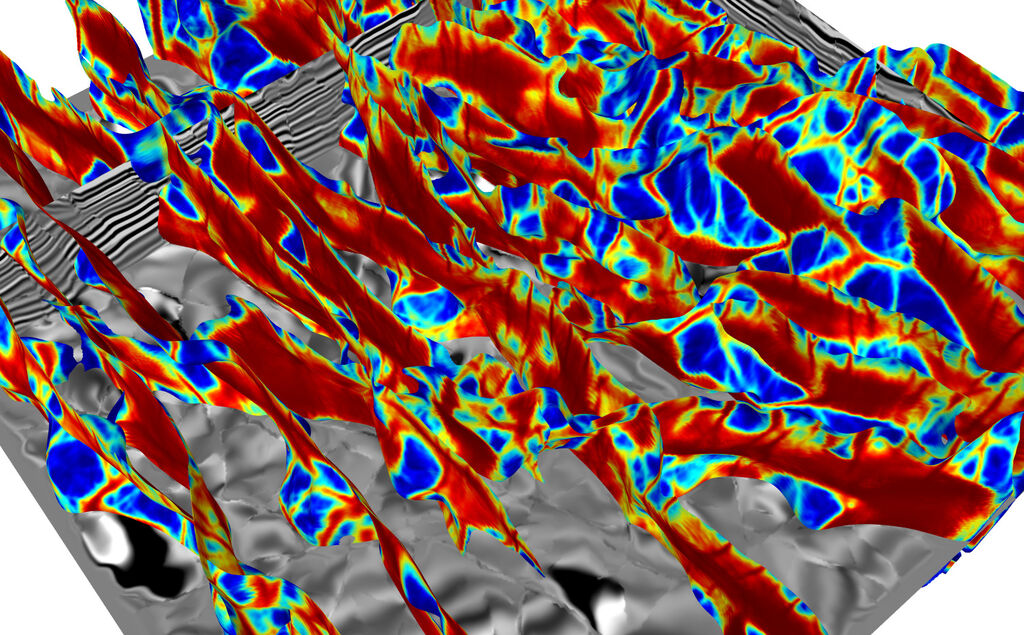}} \\
	\subfloat[]{\includegraphics[width=0.6\textwidth]{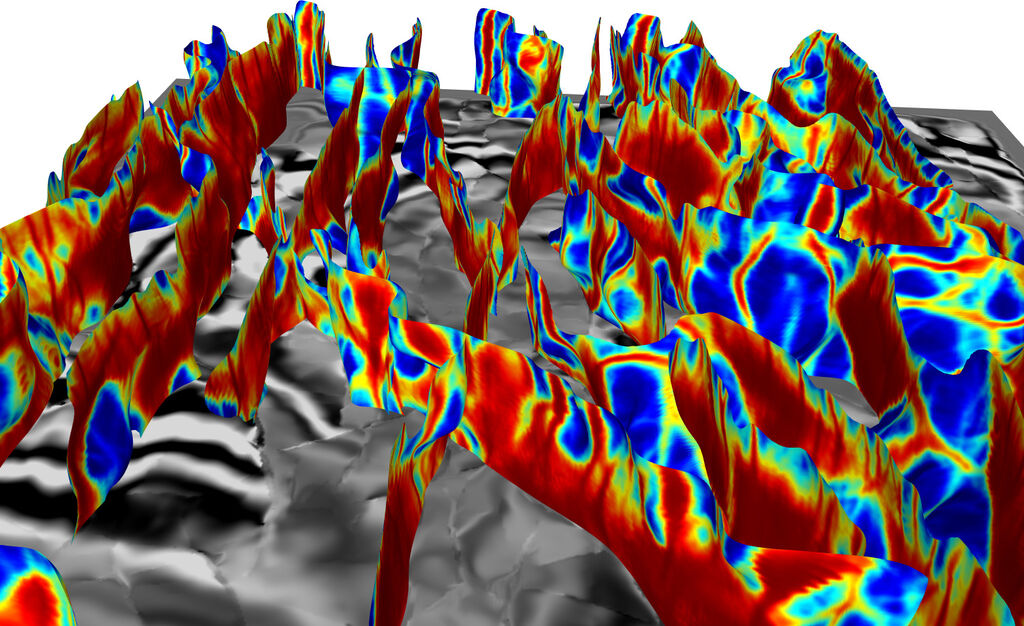}}
	\caption{Panels (a)-(c) show the detected faults from the structural image volume shown in Figure~\ref{fig:ipp_hf0_1} at three different view angles. Colors of the fault surfaces represent fault probability and are consistent with those in Figures~\ref{fig:ipp_hf0_1}-\ref{fig:ipp_hf0_3}.}
	\label{fig:fault_hf0}
\end{figure}

We further construct fault surfaces using the high-resolution image 
volume shown in Figures~\ref{fig:ipp_hf0_1}-\ref{fig:ipp_hf0_3}.  
Figure~\ref{fig:fault_hf0} show the constructed fault surfaces at three 
different view angles. These fault surfaces have relatively higher 
overall fault probability compared with those associated with the 
10-m-grid-spacing image shown in Figure~\ref{fig:fault_lf0}.  Some of 
these faults are not properly detected from the 10-m-grid-spacing image.  
We also find interlacing fault surfaces in Figure~\ref{fig:fault_hf0}, 
which further verify that the fault system at the Soda Lake geothermal 
field is fairly complex, and requires high-resolution images to reveal 
those faults. 

\begin{figure}
	\centering
	\subfloat[]{\includegraphics[width=0.45\textwidth]{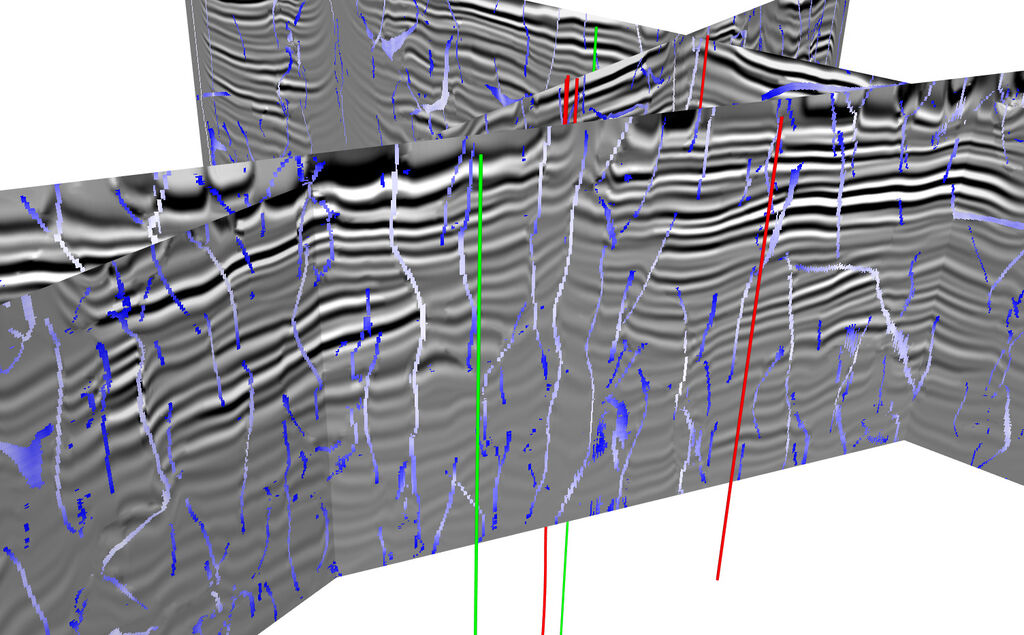}} \quad
	\subfloat[]{\includegraphics[width=0.45\textwidth]{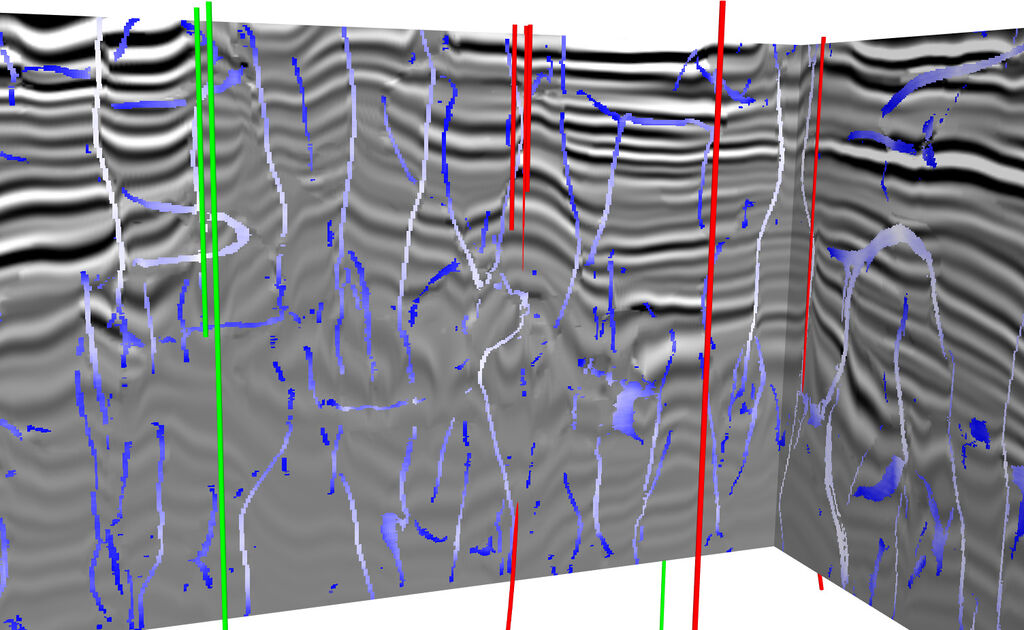}} \\
	\subfloat[]{\includegraphics[width=0.45\textwidth]{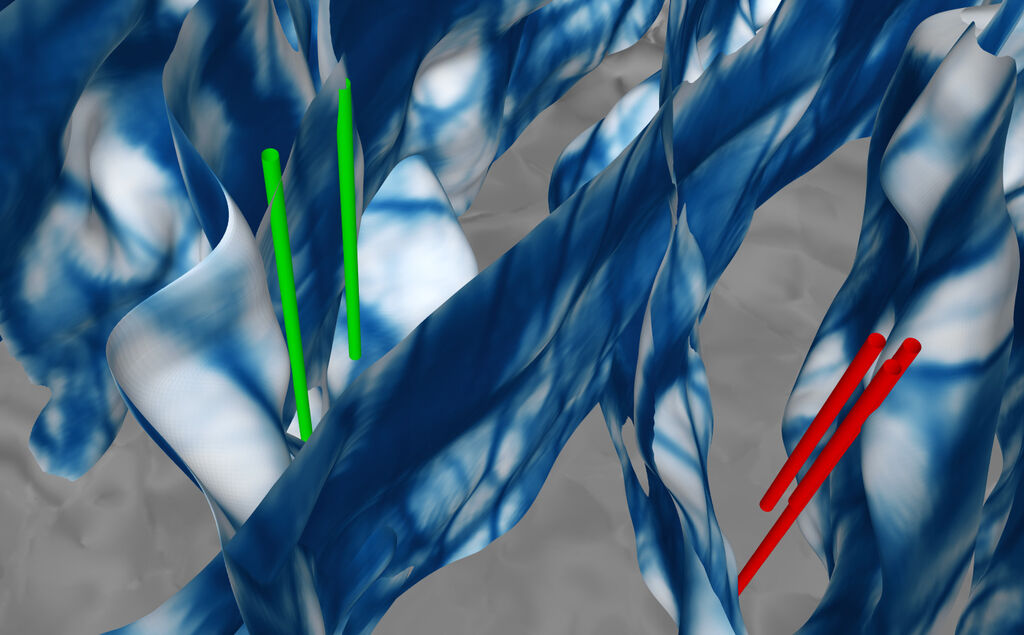}} \quad
	\subfloat[]{\includegraphics[width=0.45\textwidth]{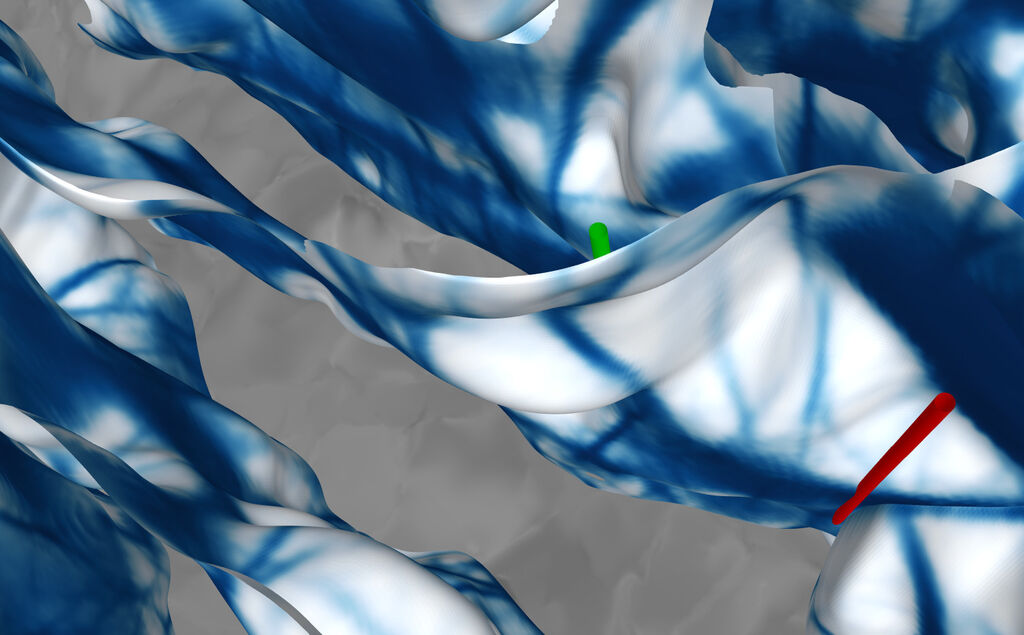}} \\
	\subfloat[]{\includegraphics[width=0.45\textwidth]{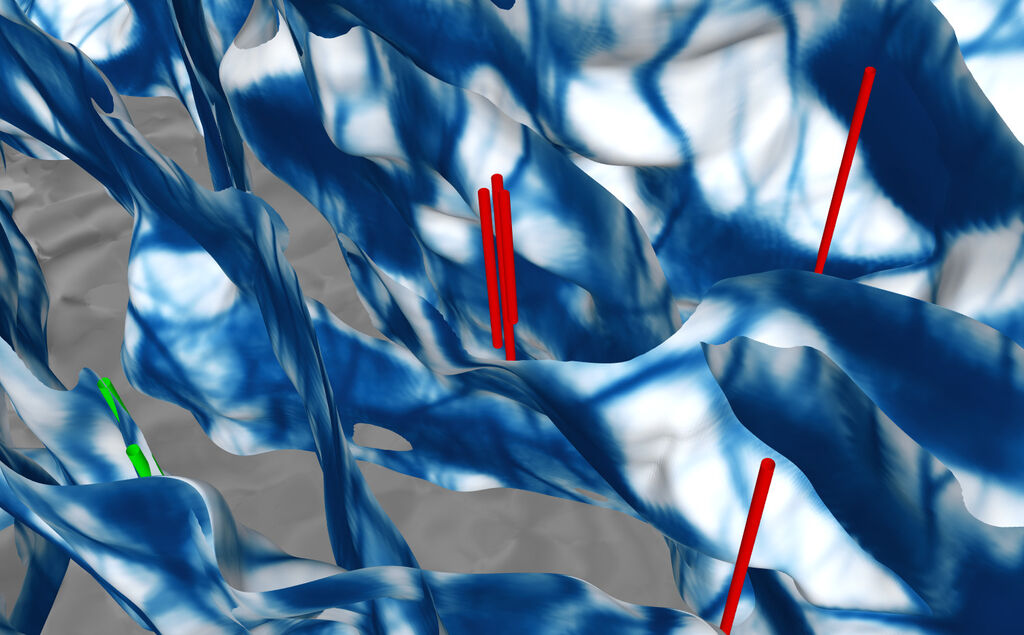}} \quad
	\subfloat[]{\includegraphics[width=0.45\textwidth]{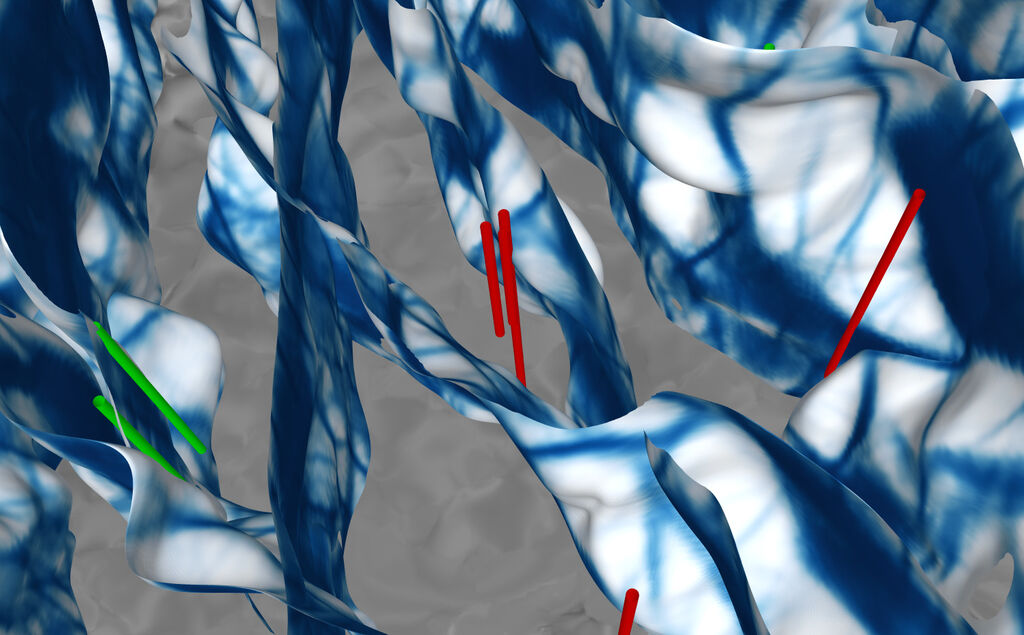}} \\
	\caption{Panels (a)-(b) show the constructed fault surfaces from 
        the structural image shown in Figure~\ref{fig:ipp_hf0_1} at two 
     different view angles, along with active injection (green) and 
  production (red) wells in this area. Panels (c)-(f) depict the 
  constructed fault surfaces with currently active injection and 
  production wells at the Soda Lake geothermal field.}
	\label{fig:well_fault_hf0}
\end{figure}

Similar to the examination on the fault-well consistency for the 
10-m-grid-spacing image, we show in Figure~\ref{fig:well_fault_hf0} that 
all currently active injection and production wells either run through or 
are very close to our detected faults. This fault-well consistency, along 
with that associated with the 10-m-grid-spacing image volume, further 
verifies the accuracy and reliability of our subsurface fault imaging at  
the Soda Lake geothermal field.

\section{Conclusions}

We have conducted a 3D, high-resolution seismic 
characterization for the Soda Lake geothermal field using full-waveform 
inversion and reverse-time migration of 3D surface seismic data.  We have 
obtained updated velocity, density, and acoustic impedance models using  
3D full-waveform inversion of the pre-processed 3D PP seismic data.  We 
have also performed high-resolution, wavefield-separation-based 
reverse-time migration to obtain high-resolution 3D structural images, 
including a 10-m-grid-spacing image up to 2.5~km in depth, and 
a high-resolution image with a vertical grid spacing of 2.5~m up to 1~km. 
We have detected faults from these image volumes and constructed 
corresponding fault surfaces, revealing the complex fault system at the 
Soda Lake geothermal field. A careful check on the consistency between 
the constructed fault surfaces and current active injection and 
production wells validate that our seismic inversion and imaging results 
and detect faults are accurate and reliable. These results can provide 
valuable information for optimizing well placement and geothermal energy 
production. Future work aims at using multi-component elastic seismic 
data to conduct isotropic and anisotropic elastic full-waveform inversion 
to reveal anisotropic characteristics of the Soda Lake geothermal field. 

\section{Acknowledgments}

This work was supported by the U.S.\ Department of Energy (DOE)  
Geothermal Technologies Office through the Los Alamos National Laboratory 
(LANL).  LANL is operated by Triad National Security, LLC, for the U.S.\ 
DOE National Nuclear Security Administration (NNSA) under Contract No.\  
89233218CNA000001.  This research used resources provided by the LANL 
Institutional Computing Program supported by the U.S.\ DOE NNSA under 
Contract No.~89233218CNA000001. We thank Xinming Wu and Benxin Chi for 
helpful discussions, and David Li for his review of the manuscript. 

\bibliographystyle{seg}
\bibliography{refs}

\end{document}